\documentclass[conference]{IEEEtran}
\IEEEoverridecommandlockouts
\usepackage{fancyhdr}
\usepackage{subcaption}
\usepackage{cite}
\usepackage{booktabs}
\usepackage{amsmath,amssymb,amsfonts}
\usepackage{algorithmic}
\usepackage{graphicx}
\usepackage{textcomp}
\usepackage{xcolor}
\usepackage{tabularx}
\usepackage{url}
\usepackage{multirow}
\usepackage[linesnumbered,ruled,vlined]{algorithm2e}
\def\BibTeX{{\rm B\kern-.05em{\sc i\kern-.025em b}\kern-.08em
    T\kern-.1667em\lower.7ex\hbox{E}\kern-.125emX}}

\newcommand{\ours}{BlockMGARD\xspace}

\begin{document}

\title{BlockMGARD: Accelerating Adaptive Scientific Data Reduction with Region-of-Interest Error Control on GPUs}


\author{
\IEEEauthorblockN{Yanliang Li}
\IEEEauthorblockA{\textit{University of Oregon}\\
Eugene, OR\\
leonli@uoregon.edu}
\and
\IEEEauthorblockN{Qian Gong}
\IEEEauthorblockA{\textit{Oak Ridge National Laboratory}\\
Oak Ridge, TN\\
gongq@ornl.gov}
\and
\IEEEauthorblockN{Qing Liu}
\IEEEauthorblockA{\textit{New Jersey Institute of Technology}\\
Newark, NJ\\
qing.liu@njit.edu}
\and
\IEEEauthorblockN{Jaemoon Lee}
\IEEEauthorblockA{\textit{Oak Ridge National Laboratory}\\
Oak Ridge, TN\\
leej8@ornl.gov}
\and
\IEEEauthorblockN{Norbert Podhorszki}
\IEEEauthorblockA{\textit{Oak Ridge National Laboratory}\\
Oak Ridge, TN\\
pnorbert@ornl.gov}
\and
\IEEEauthorblockN{Scott Klasky}
\IEEEauthorblockA{\textit{Oak Ridge National Laboratory}\\
Oak Ridge, TN\\
klasky@ornl.gov}
\and
\IEEEauthorblockN{Xin Liang}
\IEEEauthorblockA{\textit{Oregon State University}\\
Corvallis, OR\\
lianxin@oregonstate.edu}
\and
\IEEEauthorblockN{Jieyang Chen}
\IEEEauthorblockA{\textit{University of Oregon}\\
Eugene, OR\\
jieyang@uoregon.edu}

\thanks{This manuscript has been authored in part by UT-Battelle, LLC, under
contract DE-AC05-00OR22725 with the US Department of Energy (DOE).
The publisher, by accepting the article for publication, acknowledges that
the U.S. Government retains a non-exclusive, paid up, irrevocable, world-
wide license to publish or reproduce the published form of the manuscript, or allow others to do so, for U.S. Government purposes. The DOE will provide
public access to these results in accordance with the DOE Public Access Plan
(http://energy.gov/downloads/doe-public-access-plan).}
}


\maketitle
\thispagestyle{fancy}
\lhead{}
\rhead{}
\chead{}
\lfoot{\footnotesize{SC26, November 15-20, 2026, Chicago, Illinois, USA
\newline 979-8-3195-4789-7/26/\$31.00 ©2026 IEEE}}
\rfoot{}
\cfoot{}
\renewcommand{\headrulewidth}{0pt}
\renewcommand{\footrulewidth}{0pt}

\begin{abstract}

The growing scale of scientific data makes lossy compression essential for reducing data volume under controllable error. Transformation-based compressors using multilevel decomposition, such as MGARD, achieve strong compression ratios but map poorly to GPU architectures. We propose \ours, an adaptive, Region-of-Interest (ROI)-supported GPU lossy compressor, with four contributions: (1) an In-cache Block decomposition leveraging GPU on-chip memory and constant lookup tables to accelerate decomposition; (2) a hybrid hierarchy combining In-cache Block and global decomposition to balance speed and compression ratio; (3) an end-to-end pipeline with fine-grained ROI error control for feature preservation; and (4) an evaluation against state-of-the-art methods on five real-world datasets. Compared to MGARD-X, \ours achieves up to 4.2$\times$ and 9.1$\times$ higher compression and decompression throughput, and up to 8.63$\times$ higher compression ratio than uniform-tolerance baselines under ROI-aware error control. Across four GPUs, \ours achieves near-ideal linear scaling and up to 14.8$\times$ I/O cost reduction over uncompressed I/O.

\end{abstract}





\section{Introduction}





The velocity and volume of scientific data generated from large-scale simulations~\cite{chang2009compressed, ku2018full, vay2018warpx, burau2010picongpu}, observational instruments~\cite{8888528, dodson2024optimising}, and experimental facilities~\cite{Bilderback_2005} continue to grow at an unprecedented pace. Modern scientific workflows routinely produce multi-terabyte to petabyte-scale datasets, far outpacing the bandwidth and capacity growth of storage systems and I/O subsystems~\cite{dongarra2011international,io_bottleneck_2014,wan2022improving}. As a result, data movement and storage have become critical bottlenecks that increasingly limit end-to-end scientific productivity.
To cope with these constraints, scientific datasets are often truncated, downsampled, or selectively pruned before being written to storage. While these approaches reduce data size, they risk permanently discarding valuable information, including subtle features that may be essential for downstream analysis, validation, or scientific discovery~\cite{chen2019understanding}. Preserving scientific fidelity while reducing data volume has therefore emerged as a fundamental challenge in data-intensive scientific computing.

Error-controlled lossy compression has been widely recognized as a viable and effective solution to this problem~\cite{7967203,8031063,lindstrom2014fixed}. By bounding the numerical error introduced during compression, error-controlled compressors can significantly reduce data volume while maintaining user-defined accuracy guarantees. Among these approaches, transformation-based data decorrelation plays a central role. By transforming raw data into a representation that is more compressible and amenable to error control, decorrelation algorithms enable high compression ratios without violating scientific error constraints.
Representative examples include multigrid-based decomposition used in MGARD~\cite{gong2023mgard, liang2021mgard+} and near-orthogonal transform techniques employed by ZFP~\cite{lindstrom2014fixed, lindstrom2025zfp}. These methods exploit spatial or hierarchical structure in scientific data to separate coarse and fine components, thereby exposing redundancy that can be efficiently encoded. To maximize compressibility, such decorrelation algorithms typically operate over large data domains---often the entire dataset---so as to fully exploit global correlations~\cite{ainsworth2019qoi, ainsworth2019multilevel, ainsworth2018multilevel}.

Transformation-based decorrelation involves complex computation patterns with deep data dependencies across multiple stages of decomposition. These dependencies introduce substantial coordination and communication overhead, leading to low computational efficiency~\cite{8048933, chen2021accelerating}. In practice, the transformation step can dominate the overall compression pipeline, slowing down data reduction and, in some cases, even delaying I/O operations themselves~\cite{jin2022accelerating, chen2025hpdr}.
GPUs have been increasingly adopted to accelerate transformation-based decorrelation algorithms due to their massive parallelism and high memory bandwidth~\cite{chen2021accelerating, lu2023zfp, cuZFP, tian2020cusz, huang2023cuszp}. 

Despite their effectiveness, transformation-based decorrelation algorithms are poorly matched to GPU architectures. They are memory-intensive and synchronization-heavy due to deep data dependencies, offer limited parallelism, and rely on decomposition hierarchies that neither align well with the GPU memory hierarchy nor support fine-grained region-of-interest (ROI) error control needed for preserving key features in scientific data~\cite{tian2020cusz,8048933,ainsworth2019qoi,Volkov:EECS-2016-143}.

In this work, we aim to fundamentally redesign transformation-based data decorrelation to better fit GPU architectures, enabling high-performance compression with flexible and adaptive trade-offs among performance, compression ratio, and error control.
\begin{itemize}

\item We introduce a new data decomposition hierarchy, termed the \emph{In-cache Block} hierarchy, that leverages programmable on-chip memory (e.g., shared memory) on GPUs to enable fast, localized decomposition. By keeping an entire single-level decomposition within on-chip memory, this approach dramatically reduces unnecessary DRAM accesses and improves computational efficiency. Moreover, the In-cache Block hierarchy decouples the decomposition structure from input data dimensions, significantly improving performance for small or narrow datasets.

\item To enable adaptive trade-offs between performance and compressibility, we propose the first data decomposition hierarchy that combines high-throughput In-cache Block-based local decomposition with high-decorrelation-capable global decomposition, forming a \emph{hybrid hierarchy}. This hybrid approach allows fine-grained control over the balance between decomposition performance and data decorrelation efficiency, enabling the compressor to adapt to diverse application requirements and system resource constraints.

\item We build a complete hybrid-hierarchy-based lossy compression pipeline, \emph{BlockMGARD}. Compared with state-of-the-art MGARD-X compression~\cite{gong2023mgard, chen2025hpdr}, BlockMGARD significantly improves compression throughput while providing explicit trade-off control between performance and compression ratio. In addition, BlockMGARD enables fine-grained, ROI error control using the hybrid hierarchy, improving compression effectiveness while preserving scientifically important features in critical regions of the data.

\item We evaluate \ours on five real-world scientific datasets and demonstrate that our In-cache Block decomposition achieves up to 5.26$\times$ per-level throughput improvement over the original MGARD decomposition. Our hybrid hierarchy spans a continuum from up to 2.96$\times$ decomposition speedup over full global decomposition while retaining competitive compression ratios, to up to 8.1$\times$ speedup when maximizing throughput. For ROI-aware error control, \ours delivers up to 8.63$\times$ higher compression ratio than uniform-tolerance baselines at matched reconstruction fidelity, and superior visual quality at matched compression ratio. End-to-end compression evaluation further shows that, compared to the MGARD-X baseline, \ours achieves up to 4.17$\times$ and 9.13$\times$ higher compression and decompression throughput, respectively. Across four GPUs, \ours achieves near-ideal linear scaling and up to 14.8$\times$ I/O cost reduction over uncompressed I/O.
\end{itemize}
\section{Background and Related Works}

\subsection{GPU-Accelerated Lossy Compression}

Lossless compression techniques such as GZIP~\cite{deutsch1996gzip}, ZSTD~\cite{collet2015zstandard}, and FPZIP~\cite{4015488} are widely used in data management systems due to their ability to perfectly reconstruct original data. However, lossless compressors suffer from very limited compression ratios (usually less than 2$\times$) on scientific floating-point data due to the incompressible noise inherent in the low-order bits of floating-point mantissas~\cite{knorr2021ndzip}. Given the massive data volumes generated by modern simulations, lossy compression with guaranteed error bounds has been developed as a more practical solution.

Error-bounded lossy compression allows users to specify acceptable error tolerances while achieving significantly higher compression ratios (often 10:1 to 100:1 or more). Existing approaches can be broadly categorized into prediction-based methods and transformation-based methods. SZ~\cite{liang2018error}\cite{liang2022sz3}\cite{tao2017significantly}\cite{zhao2021optimizing} is a leading prediction-based compressor that performs prediction, quantization, entropy encoding, and lossless compression in sequence. SZ3 further extends this framework with a modular design allowing different predictor, quantizer, and encoder combinations to be composed for different datasets. ZFP~\cite{lindstrom2014fixed} is a representative transformation-based compressor that applies near-orthogonal block transforms followed by embedded coding to achieve fixed-rate or error-bounded compression.
MGARD~\cite{ainsworth2018multilevel}\cite{ainsworth2019multilevel}\cite{ainsworth2020multilevel}\cite{liang2021mgard+}, another transformation-based compressor, leverages multilevel decomposition based on finite element theory to provide rigorous error control on raw data and downstream Quantities of Interest (QoIs)~\cite{li2025hp}.

As demand for compression throughput to match the data generation rates of modern scientific applications grows, GPU acceleration has become necessary. cuSZ~\cite{tian2020cusz} eliminates prediction-based compression's sequential dependencies through dual quantization, enabling efficient GPU parallelization of the SZ algorithm.
cuZFP~\cite{cuZFP} provides GPU acceleration for the ZFP compressor, but is limited to fixed-rate compression mode and does not support error-bounded compression.
GPU-MGARD~\cite{chen2021accelerating} accelerates MGARD's multilevel decomposition on GPUs through optimized kernels for grid-wise, linear, and iterative processing with efficient memory access patterns.
Recent developments have further improved GPU compression performance. cuSZp~\cite{huang2023cuszp}\cite{huang2024cuszp2}\cite{huang2025gpu}\cite{zhang2025pushing} introduces progressive enhancements, including kernel fusion, optimized lossless encoding modes, and extended dimensionality support across multiple compression modes, pushing kernel throughput to hundreds of GB/s on NVIDIA GPUs.

Despite these advances in GPU compression throughput, existing approaches face inherent trade-offs between compression ratio and speed. Transformation-based methods like MGARD achieve high compression ratios through multilevel decomposition that captures global data correlations, but the intra- and cross-level data dependencies limit parallelization granularity. Prediction-based methods like cuSZp enable fine-grained parallelism but may sacrifice compression efficiency on data with strong spatial correlations. Furthermore, current GPU implementations primarily optimize global memory access patterns, leaving the potential of the on-chip memory hierarchy (shared memory and registers) largely unexplored for accelerating compression kernels. These limitations motivate the need for new algorithmic designs that can better exploit GPU memory hierarchies while providing flexible control over the compression ratio-speed trade-off.

\subsection{Multilevel Decomposition in MGARD}

MGARD~\cite{ainsworth2018multilevel}\cite{ainsworth2019multilevel}\cite{ainsworth2020multilevel}\cite{liang2021mgard+} performs decorrelation through a multilevel decomposition rooted in finite element theory. Given data on a structured grid, MGARD constructs a sequence of nested grids $\mathcal{G}_0 \subset \mathcal{G}_1 \subset \cdots \subset \mathcal{G}_{\ell}$, where each level halves the resolution along every dimension so that a dimension of size $n$ coarsens to $\lfloor n/2 \rfloor + 1$ nodes. The number of levels is therefore fixed by the grid dimensions, and every level is defined over the entire domain.

Decomposition proceeds from the finest level downward. At each level, MGARD interpolates the coarse-grid values onto the fine grid and subtracts the result from the fine-grid data, producing multilevel coefficients on the nodes that the coarse grid does not contain. These coefficients capture the detail that the coarser representation cannot express. To make the decomposition a projection in the finite element sense rather than a plain difference, MGARD then applies a correction to the coarse-grid values, computed by solving a mass matrix system derived from the underlying basis functions. For multidimensional data, interpolation and correction are applied as a sequence of one-dimensional passes, one per dimension, each shrinking the working array along the dimension it processes. The output of a full decomposition is the coarsest representation and the multilevel coefficients at every level, from which the original data can be recovered exactly by reversing the operations.

Because the coefficients at coarser levels influence a progressively larger portion of the reconstruction, MGARD assigns a level-dependent share of the error budget when quantizing them, tightening the tolerance as the level coarsens. This construction yields provable bounds on the reconstruction error in $L_\infty$ and other norms~\cite{ainsworth2019multilevel}, and extends to bounding the error of derived quantities of interest~\cite{ainsworth2019qoi,gong2022trust,banerjee2023online}.

MGARD-X~\cite{chen2021accelerating, gong2023mgard} provides a portable GPU/CPU implementation of this pipeline via the HPDR framework~\cite{chen2025hpdr, li2025hp}, organizing each level into grid-wise, linear, and iterative kernels tuned for coalesced global-memory access. Our work builds directly on MGARD-X and reuses its global decomposition in our hybrid hierarchy designs. MGARD's decomposition and our hybrid hierarchy both assume structured grids; extending error-controlled multilevel decomposition to unstructured meshes remains an active and complementary line of work~\cite{gong2024general}.

\subsection{Region-of-Interest Based Compression}

A region of interest (ROI) is a subset of the domain whose scientific value justifies preserving it at higher fidelity than the rest of the data. ROI-aware compression exploits this by allocating bits non-uniformly, increasing fidelity inside the ROI while aggressively compressing the background. What constitutes an ROI differs sharply across domains: image compression relies on semantic object detection, whereas in scientific data the ROI is typically characterized by numerical properties such as steep gradients or coherent vortical structures.
In the image compression domain, deep-learning approaches have achieved significant progress. Cai et al.\cite{cai2019end} proposed an end-to-end framework that jointly optimizes ROI prediction and coding through convolutional neural networks, achieving rate-distortion optimization via learned entropy models and hierarchical distortion losses. Jin et al.~\cite{jin2025customizable} further introduced text-controlled mask acquisition and customizable quality trade-offs between ROI and non-ROI regions. 
However, these methods optimize perceptual quality metrics rather than strict error bounds, making them unsuitable for scientific data where reconstruction must satisfy user-specified accuracy guarantees.

For scientific data compression, several works have explored spatially-varying error control to better preserve regions of scientific importance. Gong et al.~\cite{gong2022region} leverage MGARD's multilevel decomposition to automatically detect regions with rich details and establish the theoretical foundations for region-wise error control. Liang et al.~\cite{liang2020toward} preserve topological features in vector fields by computing vertex-wise adaptive error bounds. Li et al.~\cite{li2024accelerating} translate isosurface topology errors into per-block raw-data error bounds on GPUs, but require invoking an unmodified compressor once per coarse domain-decomposition block. A related line of work instead adapts compression parameters to features detected in the data rather than to a fixed region: Yakushin et al.~\cite{yakushin2020feature} couple in situ feature tracking with compression to guide reduction decisions, and Gong et al.~\cite{gong2023spatiotemporal} adapt spatiotemporal compression aggressiveness to preserve climate features such as tropical cyclones.
Together, these methods show how to decide where error tolerance should vary. But none of them integrate this decision into the compression pipeline at the kernel level.

A second line of work adds ROI support to high-performance GPU compressors, but does so at decompression time. Wang et al.~\cite{wang2025stz} proposed STZ, whose hierarchical partitioning enables random-access decompression of selected spatial regions such as 2D slices or 3D boxes without decompressing the entire dataset. Huang et al.~\cite{huang2025gpu} introduced VGC, which locates and operates on a specified ROI through block metadata and early stopping. Both provide flexible spatial access patterns, yet both apply a uniform error bound during compression: the ROI determines what is read back, not how many bits are spent. Consequently, no existing GPU compressor varies precision during compression---only after. Our block-wise decomposition makes this possible, since each block is already processed independently.

\section{Problem Statement}
Despite recent efforts on GPU acceleration, the irregular computation structure of existing multilevel decomposition is not well matched to GPU architectures, which are optimized for uniform, regular, and highly parallel workloads~\cite{Volkov:EECS-2016-143}. This mismatch introduces several critical challenges: 
(1) Decomposition algorithms are often memory-heavy, requiring repeated accesses to global memory as data are processed across multiple decomposition levels, leading to underutilized streaming multiprocessors (SMs) and poor arithmetic intensity~\cite{tian2020cusz}. 
(2) Strong data dependencies necessitate frequent communication and synchronization among GPU threads, incurring high coordination costs that do not scale well for large datasets~\cite{8048933}. 
(3) Parallelism is often fundamentally limited by data dimensionality. For example, in MGARD, the smallest dimension of the dataset can constrain the available decomposition parallelism, causing severe performance degradation for datasets with narrow dimensions~\cite{chen2021accelerating}. 
(4) Decomposition reorganizes data into hierarchical structures that are not well aligned with the GPU memory hierarchy, resulting in inefficient use of caches and shared memory~\cite{Volkov:EECS-2016-143}. 
(5) Existing decomposition hierarchies do not support fine-grained, ROI-aware error control, making it difficult to preserve critical features in selected regions while aggressively compressing less important areas.

In this work, we aim to answer the following research questions:
\begin{itemize}
    \item How can we redesign data decomposition to better suit GPU architectural characteristics?
    \item How can we enable a fine-grained trade-off between performance and compression ratio to suit diverse time and resource requirements?
    \item How can we achieve non-uniform ROI error control to preserve key scientific information in the data while maintaining high GPU compression throughput?
\end{itemize}
\section{Method}
\begin{figure}[htbp]
\centering
\includegraphics[width=1\linewidth]{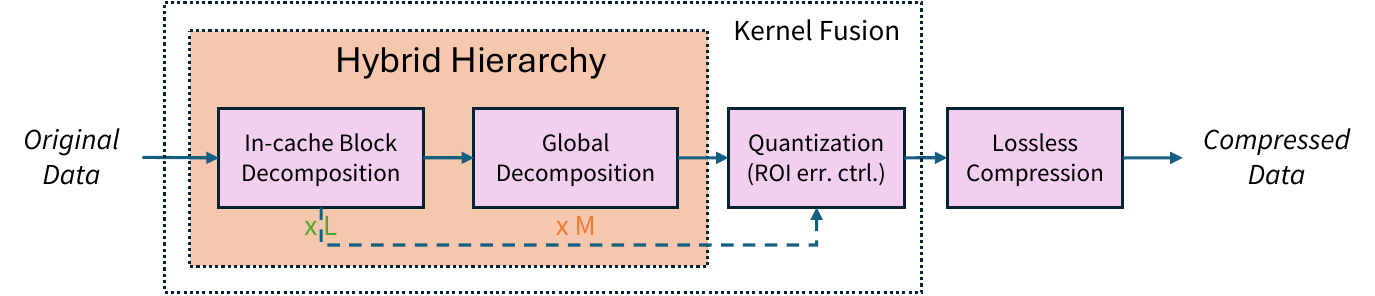}
\caption{Overview of BlockMGARD compression framework}
\label{arch}
\end{figure}

In this section, we introduce the design of BlockMGARD. As shown in Figure~\ref{arch}, BlockMGARD follows the standard transform-quantize-lossless encode pipeline; we redesign the first two stages and fuse them at the kernel level. Decomposition is structured based on a hybrid hierarchy that combines $L$ local In-cache Block levels with $M$ global levels, allowing users to trade throughput against compression ratio. We extend quantization with ROI error control at block granularity. To eliminate the global memory round trip between these two stages, we fuse decomposition and quantization into a single kernel for the local levels. We detail the In-cache Block decomposition in Section~\ref{sec:incache}, the hybrid hierarchy in Section~\ref{sec:hybrid}, the ROI error control in Section~\ref{sec:roi}, and the kernel fusion in Section~\ref{sec:fusion}.

\subsection{In-cache Block Decomposition}
\label{sec:incache}

\begin{figure}[t]
\centering
\includegraphics[width=1\linewidth]{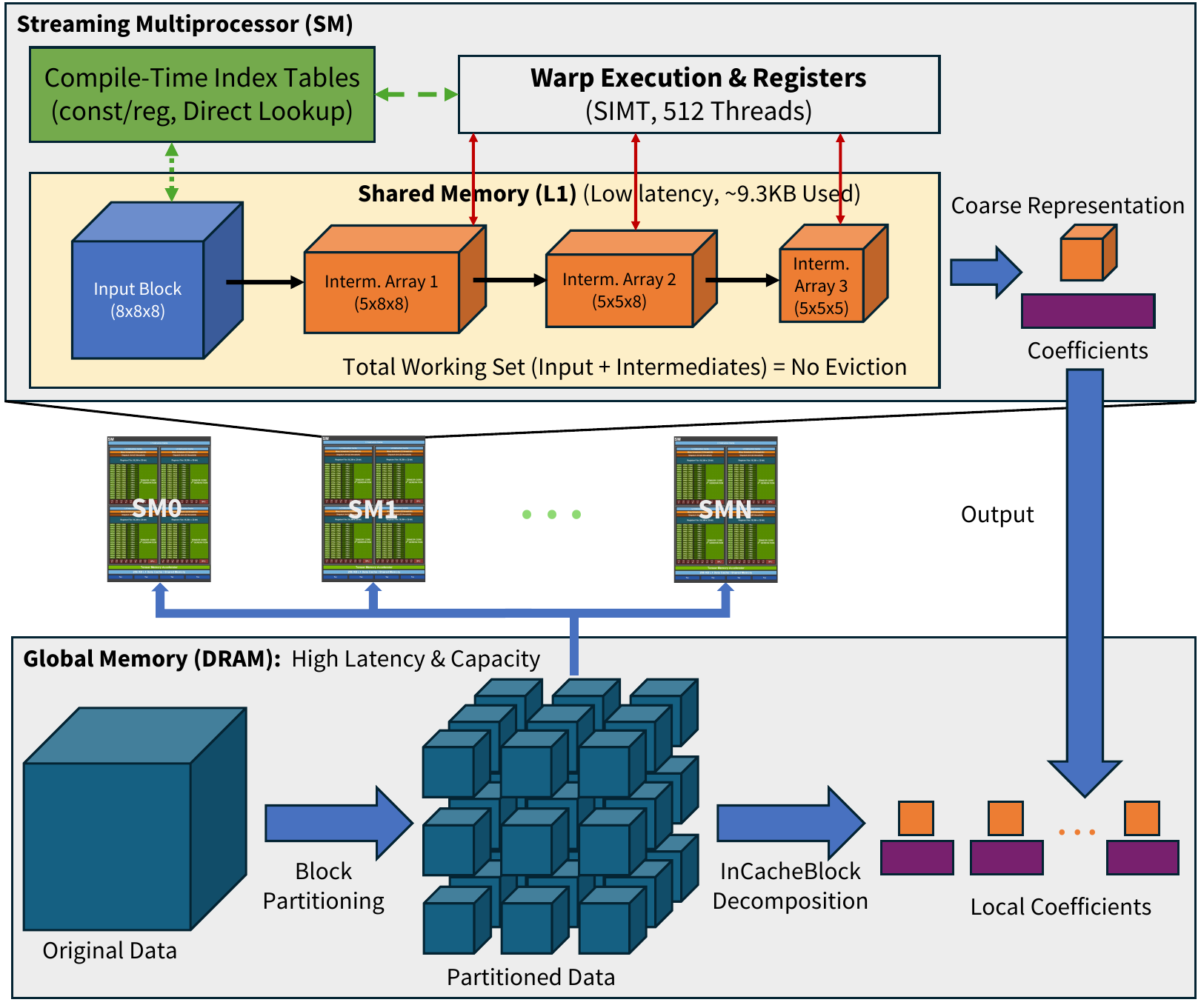}
\caption{In-cache Block decomposition mapped onto a GPU streaming multiprocessor. The entire working set---input block and intermediate arrays totaling 9.3\, KB---resides in shared memory throughout computation, with global memory accessed only for initial load and final store. Compile-time index tables eliminate runtime index arithmetic.}
\label{incacheblock}
\end{figure}

In-cache Block decomposition rests on one simple fact: on-chip memory is much faster than DRAM. Each GPU SM has its own pool of on-chip shared memory. For example, on NVIDIA Hopper GPUs, on-chip memory has an access latency of around only 29 clock cycles, versus over 650 clock cycles for accessing DRAM~\cite{luo2024benchmarking}. That gap motivates our design.
We exploit it by processing the data in small, fixed-size blocks, as shown in Figure~\ref{incacheblock}. Each block is small enough that its entire working set---the input and intermediate data---fits in shared memory. Once a block is loaded, it never leaves the chip until decomposition is done. The only global memory traffic is one coalesced load at start and one coalesced store at the end.

\paragraph{Block Size Selection}
The $8 \times 8 \times 8$ block dimensions result from co-optimization of shared memory capacity and thread parallelism. The hierarchical decomposition processes each dimension sequentially, producing intermediate arrays that must coexist in shared memory: the input block ($8 \times 8 \times 8$) and intermediate arrays of sizes $5 \times 8 \times 8$, $5 \times 5 \times 8$, and $5 \times 5 \times 5$ from successive dimensional passes. The total working set of 1157 elements occupies approximately 4.63\,KB for single-precision data and 9.26\,KB for double-precision data, fitting within a single SM's shared memory capacity (e.g., NVIDIA H100 can be configured to have up to 228\,KB shared memory) with room for lookup tables.

The $8^3 = 512$ threads organize into 16 warps, providing sufficient parallelism for latency hiding while remaining within SM thread limits. Smaller blocks such as $4^3$ yield only 64 threads (2 warps), insufficient for effective scheduling and lower decorrelation efficiency. Larger blocks such as $16^3$ require 4096 threads, exceeding SM capacity, with working sets risking shared memory overflow.

\paragraph{Compile-Time Index Resolution}
For fixed block dimensions, all indexing patterns and numerical coefficients are compile-time constants. Index mappings between input and output positions, mass matrix entries, and interpolation weights are defined as \texttt{constexpr} arrays embedded directly into the executable. This approach draws inspiration from the Flying Edges algorithm~\cite{schroeder2015flying}, which eliminates runtime case analysis through precomputed lookup tables.

During kernel execution, threads retrieve coefficients and offsets through direct table lookups indexed by thread ID, replacing expensive runtime division and modulo operations for multi-dimensional index calculation. This shifts indexing overhead to compilation, reducing per-thread instruction count and eliminating index-dependent branching that would cause warp divergence.

\paragraph{Warp Execution Uniformity}
The fixed block dimensions guarantee identical execution paths for all threads within a warp. Every In-cache Block executes the same instruction sequence without boundary-handling conditionals. Pre-computed lookup tables reinforce this uniformity: threads retrieve data through identical indexed load instructions rather than conditional arithmetic.

Shared memory on modern GPUs is organized into 32 banks with 4-byte stride. Our linear thread indexing maps consecutive threads to consecutive addresses, achieving conflict-free access for both single- and double-precision data and maximizing shared memory bandwidth.

\paragraph{Global Memory Access Pattern}
Each thread block performs exactly two global memory transactions: an initial coalesced load of the $8^3$ input block and a final coalesced store of the output. Consecutive threads access consecutive memory addresses, allowing the memory controller to combine requests into 128-byte transactions. For double-precision data, a warp's 32 threads access 256 bytes, serviced by two cache line transactions. Between these boundary accesses, all computation proceeds with zero global memory traffic.

\subsection{Hybrid Hierarchy Architecture}
\label{sec:hybrid}
While In-cache Block decomposition achieves high performance through block-wise local decomposition, its spatial partitioning has two limitations: it sacrifices global structural information that could improve compression ratio, and it may introduce boundary artifacts at block interfaces. In contrast, MGARD's global decomposition operates on the whole data domain, better preserving global features and achieving superior compression quality at the cost of performance. 

\begin{figure}[htbp]
\centering
\includegraphics[width=0.95\linewidth]{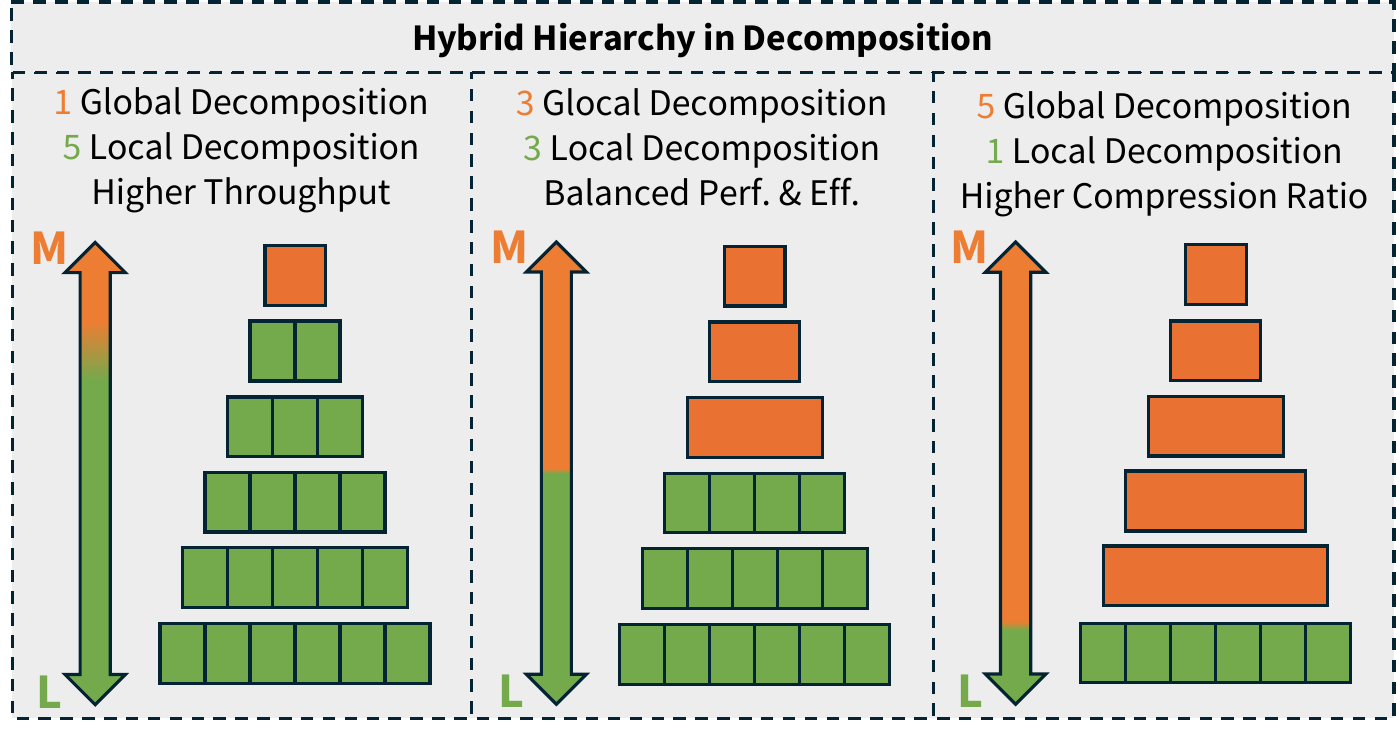}
\caption{Overview of the Hybrid Hierarchy Decomposition with different configurations. Each case represents a specific trade-off between performance and compression efficiency.}
\label{hybrid_hierarchy}
\end{figure}

We propose a hybrid hierarchy architecture that provides flexibility in balancing these objectives as shown in Figure~\ref{hybrid_hierarchy}: the first $L$ levels use In-cache Block local decomposition for maximum GPU efficiency, followed by $M$ levels of MGARD's global decomposition to reach the desired total hierarchy depth $L+M$. This design allows users to control the speed-quality trade-off through the parameter $L$. Larger $L$ values prioritize speed by performing more decomposition locally, reducing global memory operations and achieving faster compression. Smaller $L$ values leverage more global processing to achieve better compression quality with fewer artifacts, at the cost of increased runtime. The parameter $M$ determines the final coarseness of the hierarchical representation, affecting the achievable compression ratio.


\begin{figure}[t]
\centering
\includegraphics[width=1\linewidth]{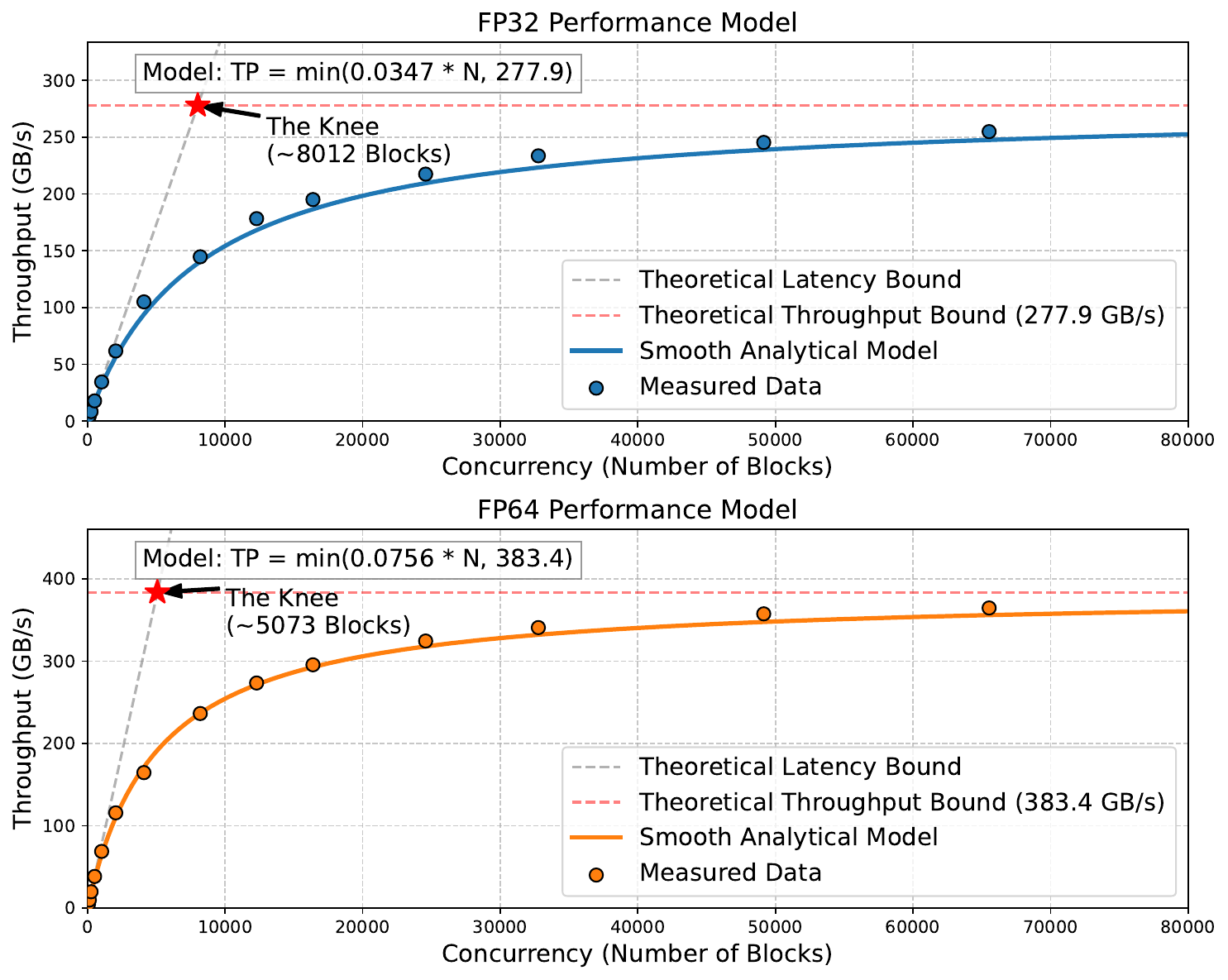}
\caption{Throughput modeling of the decomposition kernel on H100. The analytical model (solid curves) accurately predicts the measured performance (dots) by combining the theoretical latency bound and maximum throughput bound. The "knee" indicates the threshold where hardware utilization saturates.}
\label{throughput_model}
\end{figure}

To determine the optimal parameter value for $L$ and $M$ that maximize the compression ratio while keeping the total cost under a time budget, we first aim to understand the performance characteristics of In-cache Block kernel via profiling. As shown in Figure~\ref{throughput_model}, throughput scaling exhibits two distinct regimes impacted by different hardware bottlenecks.
In the first regime, when the number of $8\times8\times8$ blocks is small, throughput scales approximately linearly with block count. Since each block is mapped to a single CUDA thread block of 512 threads and each SM supports a maximum of 2048 concurrent threads, only 4 blocks can execute simultaneously per SM. H100 provides 132 SMs, yielding an SM occupancy saturation point of $132 \times 4 = 528$ blocks. Below this point, each additional block directly occupies an available SM slot, and throughput grows proportionally.
In the second regime, beyond 528 blocks, all SM slots are occupied, yet throughput continues to grow substantially. Although each In-cache Block performs its decomposition and recomposition entirely within shared memory, it still needs one coalesced load and one coalesced store per block. H100's asynchronous memory pipeline requires a large number of concurrent in-flight requests to reach full utilization. As block count increases, the GPU scheduler overlaps load and store operations from queued blocks with the on-chip computation of active blocks, progressively filling the memory pipeline until peak throughput is reached.

We model this throughput behavior using the two-regime framework proposed by Volkov~\cite{Volkov:EECS-2016-143}, which separates performance into a latency-bound regime where throughput scales linearly with concurrency, and a throughput-bound regime where performance saturates at the hardware limit. However, a strict piecewise linear model ($min$ function) assumes perfect overlap of execution and memory delays, which rarely holds in practice due to structural hazards and resource contention. 
Therefore, incorporating the principles from queueing theory and Little's Law~\cite{little1961proof}, we derive a smooth analytical expression (Equation~\ref{eq:throughput_model}) based on an additive execution time model. By superimposing the ideal latency-bound execution time and the asymptotic throughput-bound congestion-time, we formulate the effective throughput as the reciprocal sum of these limits. Unlike a discrete piecewise linear model, this formulation captures the diminishing returns in throughput as the system approaches architectural saturation.
\begin{equation}
    \label{eq:throughput_model}
    \text{TP}(N) = \frac{N}{\dfrac{1}{Slope} + \dfrac{N}{\text{TP}_{\text{peak}}}}
\end{equation}
where $N$ is the number of concurrent blocks, $Slope$ is the per-block throughput slope estimated via linear regression over the latency-bound regime ($N<528$), and $\text{TP}_{\text{peak}}$ is the measured peak throughput. Our analytical model identifies the \textit{knee point}---where the marginal throughput gains become negligible---at approximately 8{,}012 blocks for FP32 and 5{,}073 blocks for FP64.

The knee point serves as a performance reference: any configuration where the block count exceeds the knee point guarantees that the In-cache Block kernel operates in the throughput-saturated regime. Given this insight, we propose an automatic parameter selection strategy as shown in Algorithm~\ref{alg:hyperparam_select}. 
We first enumerate all valid (L, M) pairs via performance modeling and prediction, and filter those satisfying the time budget $T_{bgt}$. 
Each prediction function is a simple formula fit to profiling data on the target GPU (e.g., H100), not a full simulation. \textsc{PredictDecompTime} adds up two pieces: local levels use the throughput model from Equation~\ref{eq:throughput_model}, and global levels use a separate linear fit, since MGARD's global decomposition doesn't show the same saturation behavior as In-cache Block decomposition. \textsc{PredictQuantTime} works the same way for local and global quantization---each is $max(slope\times N, T_{latency})$, capturing both the per-element cost and a fixed latency floor we saw in practice. \textsc{PredictLosslessTime} breaks Huffman coding~\cite{tian2021huffman} into its stages---outlier handling, histogramming, codebook building, encoding, and deflate---and fits each one separately, since they don't all scale the same way with input size. We add up all three predictions and check them against $T_{bgt}$.
Finally, we select the configuration that lexicographically maximizes $(M, -L)$---prioritizing more global decomposition levels and fewer local levels to maximize the decorrelation efficiency.

\begin{algorithm}[t]
\caption{Automatic Parameter Selection for Hybrid Hierarchy}
\label{alg:hyperparam_select}
\KwIn{Data type \texttt{dtype}, shape \texttt{S}, time budget $T_{bgt}$ (ms)}
\KwOut{Recommended configuration $(L^*, M^*)$}
\texttt{candidates} $\gets$ \textsc{GenCandidates}(\texttt{S}) \tcp*{enumerate all valid $(L, M)$ pairs}
$\mathcal{F} \gets \emptyset$ \\
\For{$(L, M) \in \texttt{candidates}$}{
    $t_d \gets$ \textsc{PredictDecompTime}(\texttt{dtype}, \texttt{S}, $L$, $M$) \\
    $t_q \gets$ \textsc{PredictQuantTime}(\texttt{dtype}, \texttt{S}, $L$, $M$) \\
    $t_e \gets$ \textsc{PredictLosslessTime}(\texttt{dtype}, \texttt{S}, $L$, $M$) \\
    \If{$t_d + t_q + t_e \leq T_{bgt}$}{
        $\mathcal{F} \gets \mathcal{F} \cup \{(L, M)\}$
    }
}
\Return $\arg\max_{(L,M)\in\mathcal{F}}\ (M,\ {-L})$ \tcp*{maximize $M$, break ties by minimize $L$}
\end{algorithm}

\subsection{Region-of-Interest Error Control}
\label{sec:roi}

Scientific datasets often contain regions of varying importance, where critical features require stricter error control than background areas. We propose an ROI-based compression scheme that supports spatially-varying error tolerances at block granularity, naturally integrating with our In-cache Block decomposition architecture.

\begin{figure}[htbp]
\centering
\begin{subfigure}[t]{0.542\columnwidth}
\includegraphics[width=\linewidth]{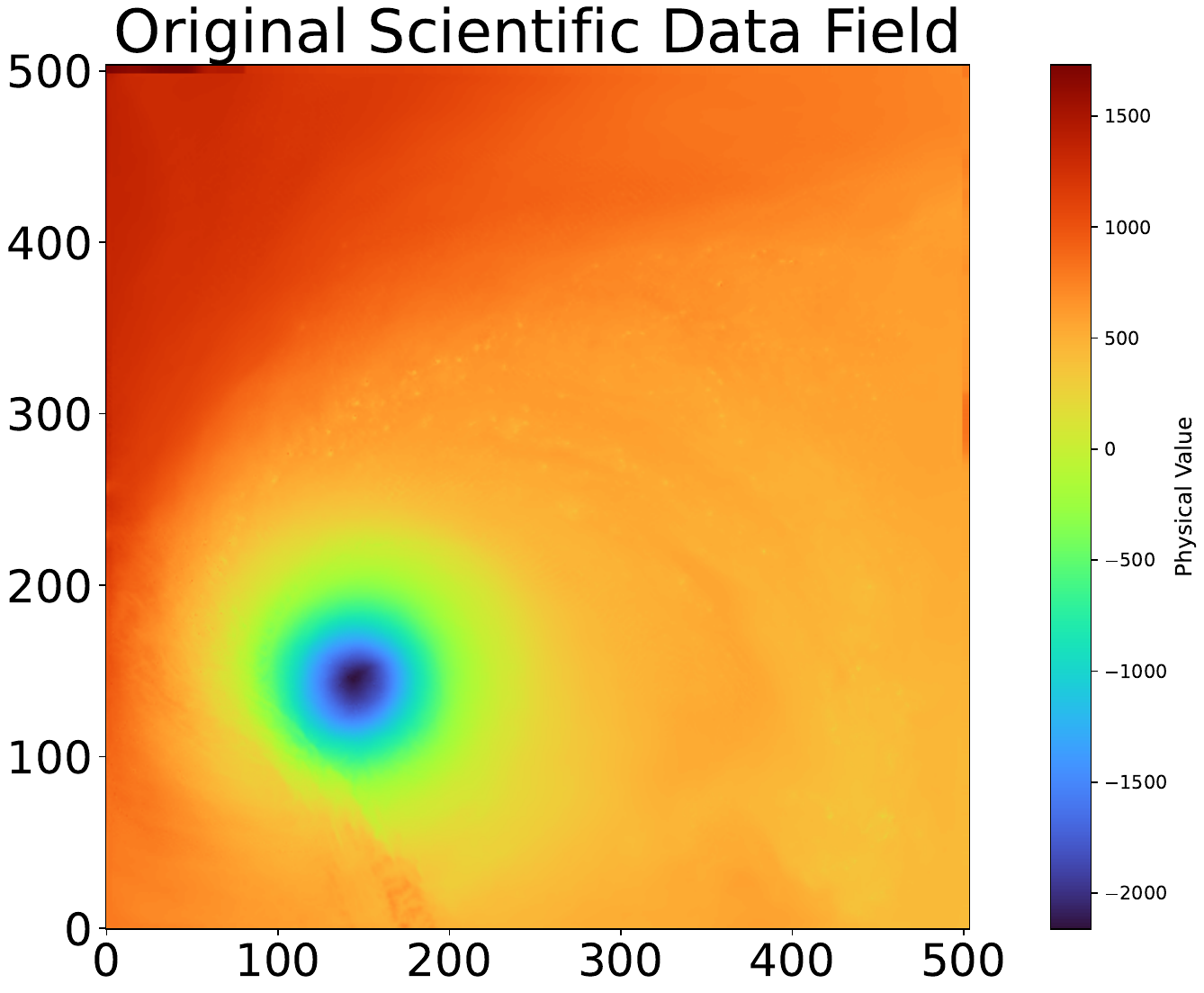}
\vspace*{-1.5em}
\caption{Dataset Visualization}
\end{subfigure}%
\hfill%
\begin{subfigure}[t]{0.44\columnwidth}
\includegraphics[width=\linewidth]{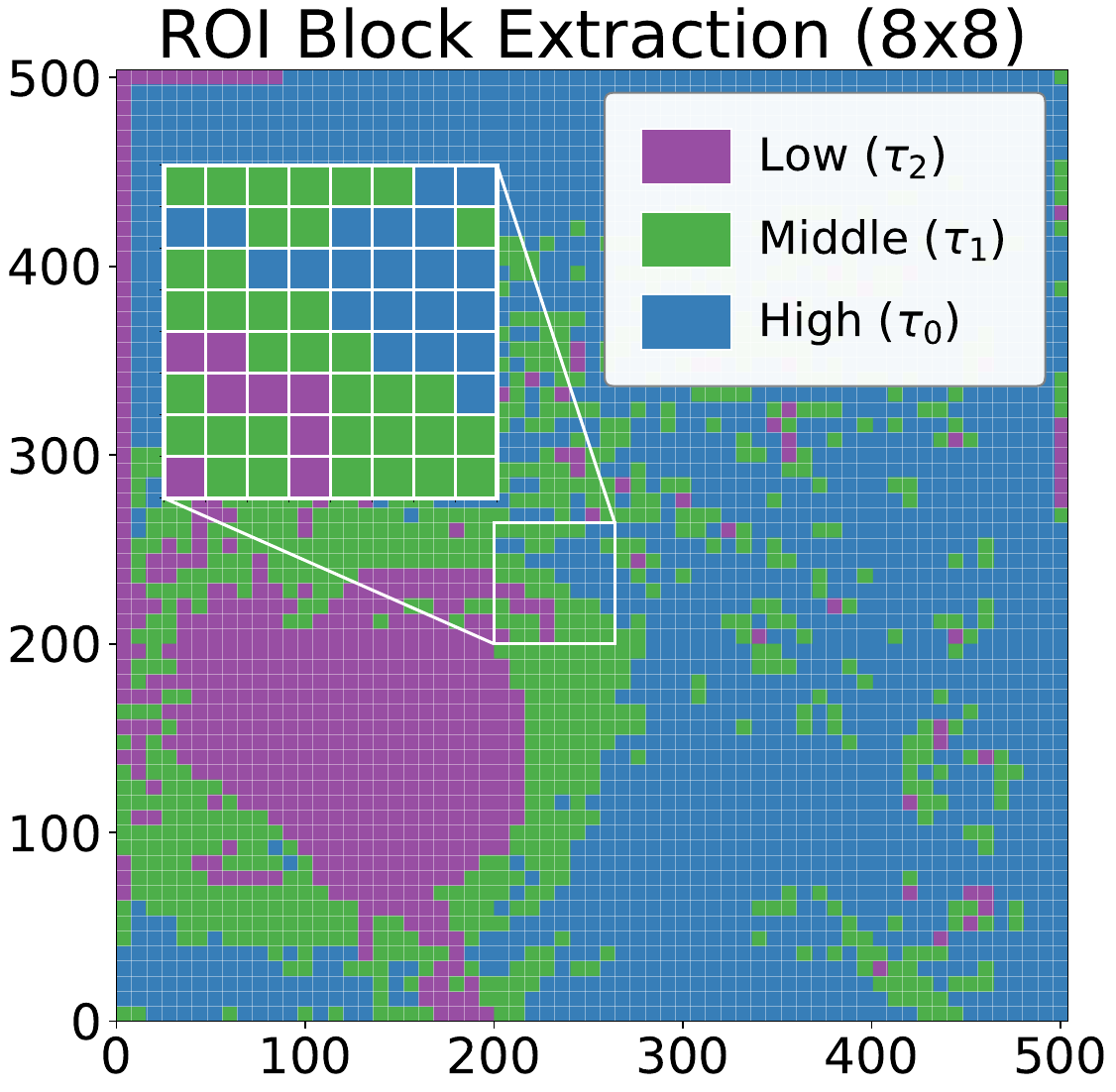}
\vspace*{-1.5em}
\caption{ROI Extraction}
\end{subfigure}
\caption{ROI extraction example on a scientific dataset. (a) Original data field showing a prominent central feature. (b) Block-level ROI assignment discretized to $8 \times 8$ granularity: strict tolerance $\tau_2$ (magenta) covers the central feature, normal tolerance $\tau_1$ (green) captures the transition region, and background tolerance $\tau_0$ (blue) applies elsewhere.}
\label{fig:roi}
\vspace*{-1em}
\end{figure}

\paragraph{Block-Wise Tolerance Assignment}
Our ROI implementation assigns tolerance values at the block level, aligning with the In-cache Block decomposition structure. Let $B = \{B_1, B_2, \ldots, B_K\}$ denote the set of blocks partitioning the domain, where $K$ is the total block count. Each block $B_i$ is assigned a tolerance $\tau_i$ that specifies the maximum allowable error for data points within that block.

Users specify ROI through a tolerance map that can be defined programmatically (e.g., based on gradient magnitude or feature detection) or through explicit region specification. The tolerance map is discretized to block resolution: if a user-defined ROI region partially overlaps a block, that block receives the stricter ROI tolerance to ensure complete coverage of the region of interest.

Figure~\ref{fig:roi} illustrates this process on a 2D slice from the pressure field (Pf48) of the Hurricane ISABEL dataset. The original data field (a) contains a central low-pressure feature surrounded by gradient regions. In (b), we demonstrate ROI extraction using a value-based criterion: data points within the lowest 15\% of the value range are assigned strict tolerance $\tau_2$, those within 15--40\% receive normal tolerance $\tau_1$, and the remainder is treated as background with tolerance $\tau_0$. This value-based approach is one example; users may alternatively specify ROI regions through gradient magnitude thresholds, explicit spatial masks, or application-specific feature detectors.

We support multiple tolerance tiers: strict ROI ($\tau_2$) for critical features requiring maximum preservation, normal ROI ($\tau_1$) for important regions with moderate precision requirements, and background ($\tau_0$) for non-critical regions allowing aggressive compression, satisfying $\tau_2 < \tau_1 < \tau_0$.

\paragraph{Error Propagation across Tolerance Boundaries}
The primary challenge in our ROI-based compression is maintaining error bounds when blocks with different tolerances contribute to the same coarse-level block through the hybrid decomposition hierarchy. 

\begin{figure}[htbp]
\centering
\includegraphics[width=0.95\linewidth]{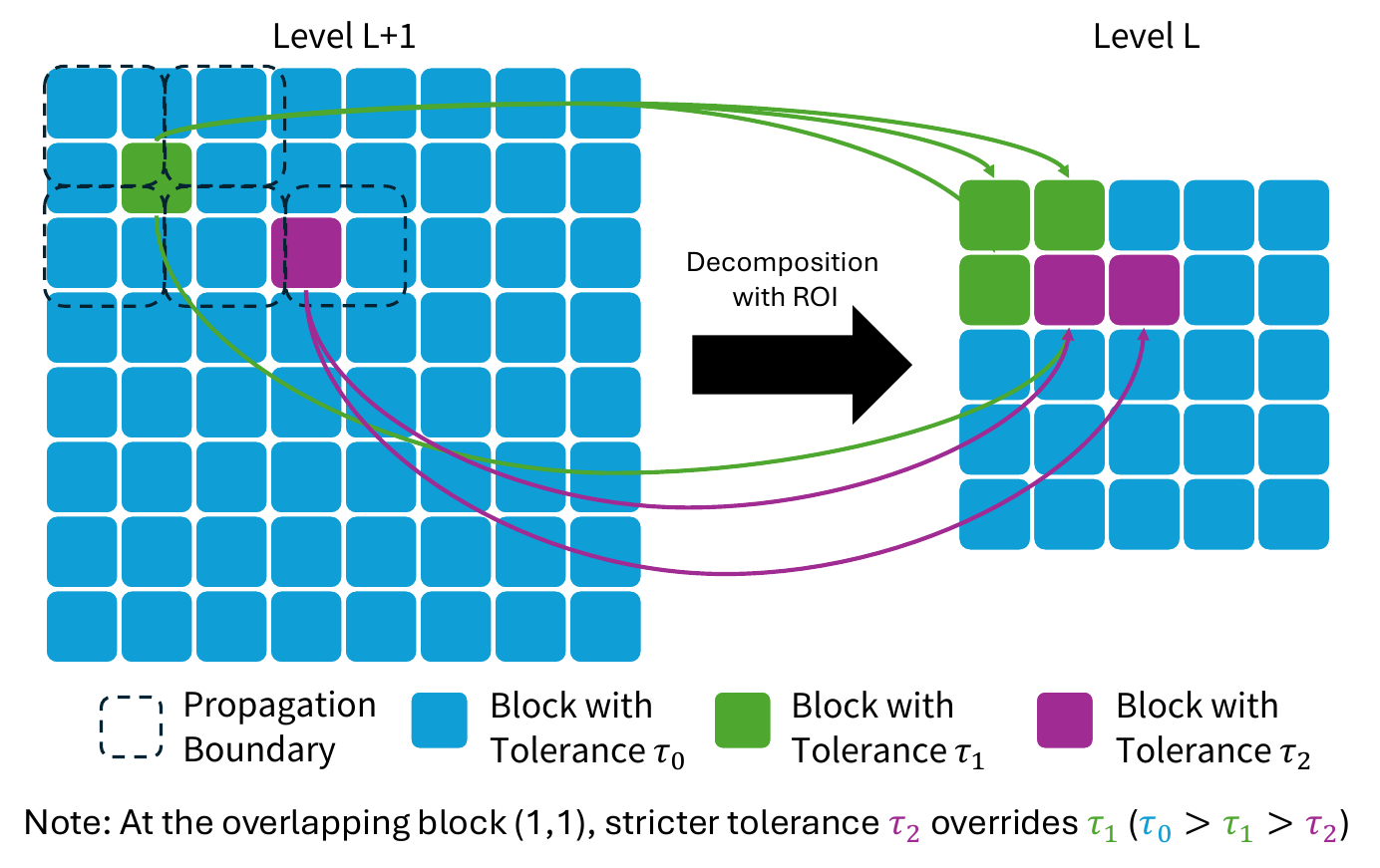}
\caption{ROI tolerance propagation from fine level (left) to coarse level (right). At Level $L+1$, three adjacent blocks have tolerances $\tau_0$, $\tau_1$, and $\tau_2$. After decomposition, the coarse outputs are assembled into new blocks at Level $L$. Block $(1,1)$ receives contributions from all three tolerance regions; conservative propagation assigns $\tau_2$ (the strictest) to guarantee $L_\infty$ bounds.}
\label{roi_propagation}
\end{figure}

The propagation issue arises from the spatial reorganization between decomposition levels. At Level $L+1$, each $8 \times 8 \times 8$ block decomposes to a $5 \times 5 \times 5$ coarse representation with detail coefficients. When proceeding to Level $L$, these $5 \times 5 \times 5$ outputs must be assembled into new $8 \times 8 \times 8$ blocks for continued decomposition. Due to the size mismatch, a single block at Level $L$ may aggregate coarse elements originating from multiple Level $L+1$ blocks with different assigned tolerances.

Figure~\ref{roi_propagation} illustrates this scenario. At Level $L+1$, three adjacent blocks have tolerances $\tau_0$ (background), $\tau_1$ (normal ROI), and $\tau_2$ (strict ROI), where $\tau_0 > \tau_1 > \tau_2$. The dashed boundary indicates the propagation region where tolerance mixing occurs. After decomposition, their $5 \times 5 \times 5$ coarse outputs are spatially adjacent. When forming the $8 \times 8 \times 8$ blocks at Level $L$, block $(1,1)$ contains tolerance contributions derived from all three tolerance regions.

If block $(1,1)$ were quantized using the more permissive tolerance $\tau_0$ or $\tau_1$, the resulting quantization error would propagate through recomposition. Since this coarse data ultimately reconstructs fine-level values belonging to the strict ROI region, the reconstructed values would violate the required $L_\infty$ bound $\tau_2$.

\begin{table*}[t]
\caption{Datasets used for evaluation}
\label{dataset}
\vspace{-0.5em}
\centering
\begin{tabular}{|c|c|c|c|c|}
\hline
Dataset     & Variable & Dimensions & Type & Size  \\ \hline
NYX &  temperature, velocity\_x/y/z  & $512 \times 512 \times 512$  &  FP32  &  2 GB\\ \hline
Hurricane ISABEL  & Pf48, Uf48, Vf48, Wf48 & $100 \times 500 \times 500$        & FP32        & 380 MB \\ \hline
SCALE-LETKF  & PRES, T, U, V & $98 \times 1200 \times 1200$        & FP32        & 2.1 GB  \\ \hline
Miranda  & density, diffusivity, pressure, velocityz & $256 \times 384 \times 384$        & FP64        & 1.1 GB \\ \hline
S3D  & CH4, CO2, H2O, O2 & $500 \times 500 \times 500$        & FP64        & 4 GB \\ \hline
\end{tabular}
\end{table*}

\paragraph{Conservative Tolerance Propagation Strategy}
To guarantee $L_\infty$ error bounds across all tolerance regions, we adopt a conservative propagation rule: each coarse-level block inherits the minimum (strictest) tolerance among all fine-level blocks that contribute to it. Formally, let $\mathcal{C}(B_j^L)$ denote the set of Level $L+1$ blocks whose coarse outputs contribute to block $B_j^L$ at Level $L$. The propagated tolerance is:
\begin{equation}
    \tau_j^L = \min_{B_i^{L+1} \in \mathcal{C}(B_j^L)} \tau_i^{L+1}
\end{equation}

This propagation is applied recursively through all $L$ levels of In-cache Block decomposition. A block at level $\ell$ inherits the strictest tolerance from its contributing blocks at level $\ell+1$, which in turn inherited from level $\ell+2$, and so on. Consequently, strict ROI tolerances propagate outward through the hierarchy, creating an expanding ``influence region'' at coarser levels.

The conservative strategy guarantees correctness but introduces overhead: blocks that contain no strict-ROI data in their original fine-level region may still receive strict tolerances due to propagation. This results in more bits allocated to these blocks than strictly necessary. However, since ROI regions are typically spatially concentrated, the propagation influence remains localized. The overhead grows with the number of local decomposition levels $L$, but for moderate $L$ values, the propagation region extends only a limited distance beyond the original ROI boundary.

When global MGARD decomposition levels ($M > 0$) follow the local levels, the error budget for global processing derives from the strictest tolerance at the final local level: $\tau_{global} = \min_i \tau_i^1$, where $\tau_i^1$ denotes the propagated tolerance of block $i$ at Level 1 (the last local level). This budget is then distributed equally across the $M$ global levels, allocating $\tau_{global}/M$ to each level. This uniform allocation maintains the mathematical guarantees of MGARD's error control framework without requiring modifications to its global processing pipeline.

\subsection{Pipeline Stage Fusion}
\label{sec:fusion}

The unfused pipeline runs In-cache Block decomposition and quantization as two separate kernels. Decomposition writes each level's coefficients to global memory in full precision. Quantization reads them back, applies the level's tolerance---a scalar quantizer, or a per-block reciprocal quantizer derived from the ROI tolerance map---and writes out quantized symbols. For each $8\times8\times8$ block, this means a full-precision round trip through global memory for detail coefficients that are only ever used once.

We fuse quantization directly into the In-cache Block kernel to avoid this round trip. 125 of the 512 threads still write out the $5\times5\times5$ coarse output in full precision, since the next level (or the global hierarchy, at the final local level) needs it. The other 387 threads compute detail coefficients instead. While those values are still sitting in shared memory, each thread applies the block's quantizer on the spot and writes the quantized symbol straight to global memory. Only the quantized index ever touches HBM, and one kernel launch per local level disappears along with it. The fused kernel also reads its input with bounds checks instead of requiring pre-padded input, so the memset-and-copy step used to pad each level's input goes away too---out-of-range positions are simply zero-filled in-kernel. This fusion applies to the $L$ local levels only; the $M$ global levels have no per-block resident working set to fuse against and keep running as separate kernels.

\section{Experimental Evaluation}

\subsection{Experimental Setup}

\paragraph{Evaluation Platform}
We conduct our experimental evaluation using the Talapas cluster~\cite{racs} from the University of Oregon, using a compute node equipped with 4 NVIDIA H100 GPUs (80 GB HBM3, 132 SMs, SM 9.0), CUDA Toolkit 12.4, and two 24-core Intel Xeon CPUs with 1,024 GB memory. 

\paragraph{Comparison Baselines and Error Bounds}

We evaluate compression performance against three GPU-accelerated 
compressors: MGARD-X\cite{chen2021accelerating}, 
cuZFP\cite{cuZFP}, and VGC\cite{huang2025gpu}. 

Our evaluation is based on the $L_\infty$-norm relative error bound: we set REL error bounds of $1e^{-2}$, $1e^{-4}$, and $1e^{-6}$, and every reconstructed value satisfies $|u-u'|_\infty < tol \times |u|_\infty$. VGC instead defines its error bound point-wise as a fraction of the data range ($d_{max}-d_{min}$), so we convert our $L_\infty$-norm relative tolerance into the corresponding absolute error bound before passing it to VGC. cuZFP only supports fixed-rate mode, so for each target error bound we manually measure the fixed rate that achieves the equivalent error and use that rate in our evaluation.

\paragraph{Datasets}

We evaluate on five real-world HPC simulation datasets from the Scientific Data Reduction Benchmarks suite\cite{sdrbench}\cite{zhao2020sdrbench}: the NYX cosmology simulation\cite{almgren2013nyx}, the Hurricane ISABEL weather simulation \cite{hurricane}, the SCALE-LETKF climate simulation\cite{hunt2007efficient}, the Miranda hydrodynamics simulation\cite{cook2004mixing}, and the S3D combustion simulation\cite{chen2009terascale}. Table~\ref{dataset} presents dataset details and the variables used in the evaluation. To ensure a fair comparison, we use four variables per dataset and report the average of the measured results across all variables.



\subsection{In-cache Block Decomposition}

We first evaluate the performance of In-cache Block decomposition by comparing it against the original global decomposition under controlled conditions. To accurately compare the efficiency of each approach, we measure only kernel execution time, excluding data transfer overhead between host and device memory. For a fair comparison, we apply a single level of decomposition to four variables for each dataset and report the average throughput.


\begin{figure}[t]
\centering
\includegraphics[width=1\linewidth]{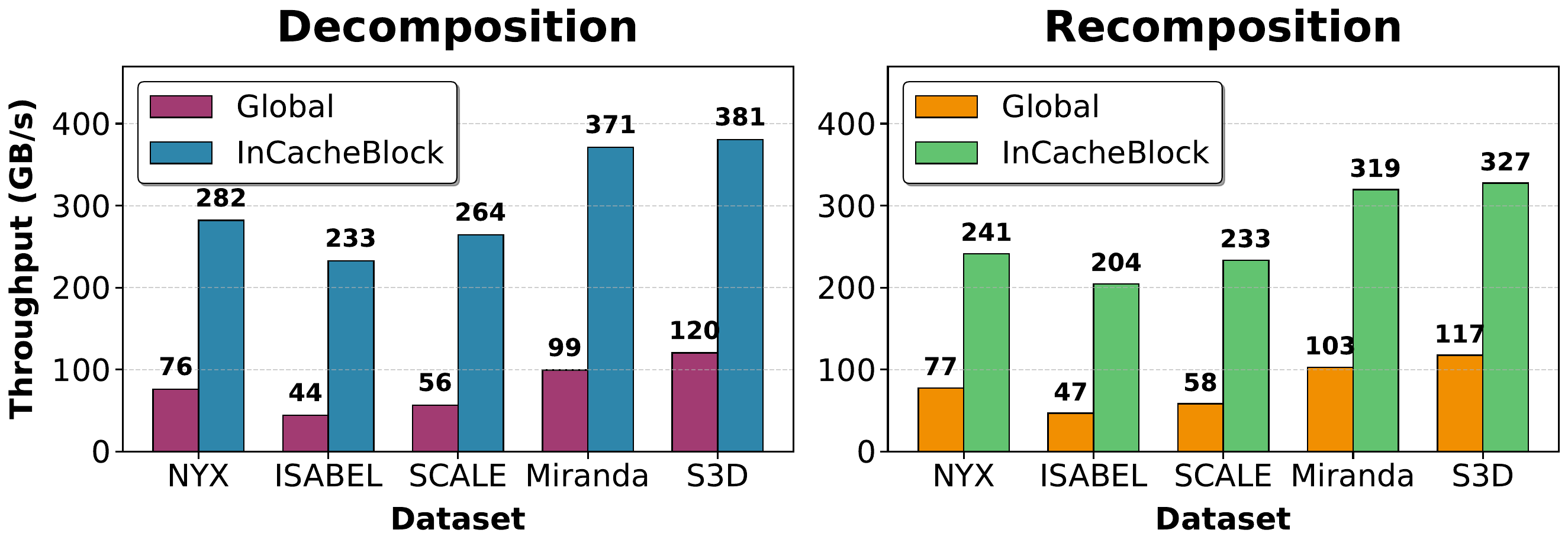}
\caption{Comparing performance of decomposing and recomposing one level of data using the original global hierarchy method vs. In-cache Block}
\label{exp_incacheblock_tp}
\end{figure}

\begin{figure*}[htbp]
\centering
\includegraphics[width=1\linewidth]{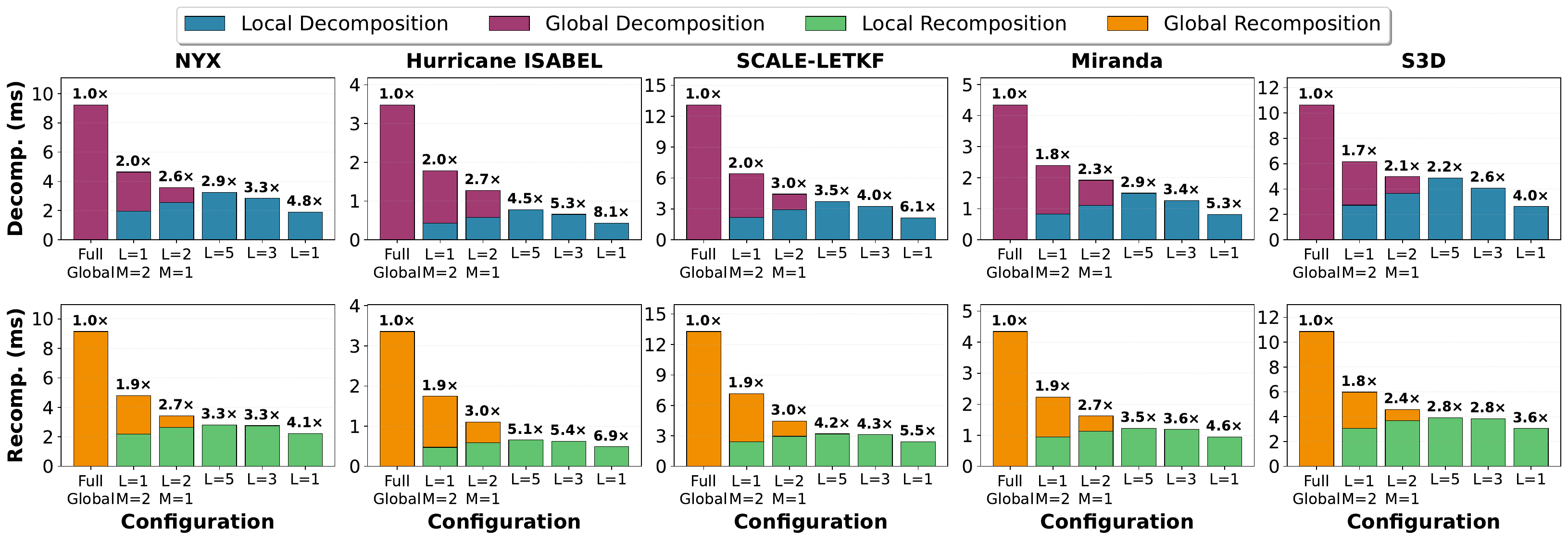}
\caption{Comparing decomposition and recomposition performance between global, hybrid, and local hierarchy.}
\label{exp_hybrid_breakdown}
\end{figure*}

\begin{figure*}[t]

    \centering
    \includegraphics[width=1\linewidth]{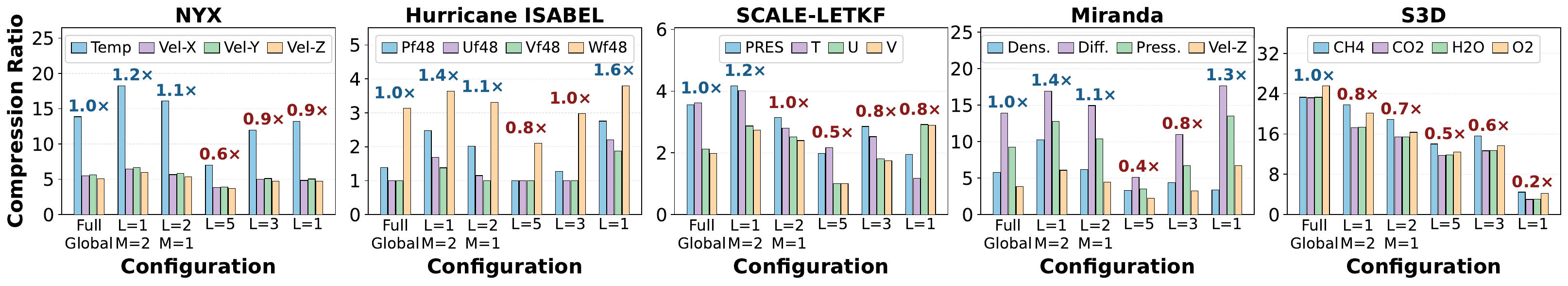}
    \caption{Comparing compression ratio between between global, hybrid, and local hierarchy}
    \label{exp_hybrid_decomposition_cr}
\end{figure*}

Figure~\ref{exp_incacheblock_tp} presents the throughput comparison across all five datasets. In-cache Block achieves 233--381 GB/s for decomposition compared to 44--120 GB/s for global decomposition, bringing a speedup of 3.16$\times$--5.26$\times$ (avg. 4.1$\times$). For recomposition, In-cache Block achieves 204--327 GB/s versus 47--117 GB/s, with a speedup of 2.79$\times$--4.40$\times$ (avg. 3.5$\times$). The consistent speedup across all datasets validates that confining the working set to shared memory effectively reduces global memory traffic. The speedup is particularly pronounced on datasets with narrow dimensions, such as Hurricane-ISABEL and SCALE-LETKF. In global decomposition, the smallest dimension of the dataset can severely constrain the available decomposition parallelism, leading to substantial performance degradation due to under-utilized hardware resources and heavy global memory traffic. In contrast, our In-cache Block decomposition effectively mitigates this bottleneck by confining the working set to shared memory, which decouples the decomposition process from global memory hierarchy constraints and ensures high throughput even when dataset dimensions are restrictive.

\subsection{Hybrid Hierarchy Decomposition}

Figure~\ref{exp_hybrid_breakdown} compares decomposition and recomposition kernel execution times across five datasets. The Full Global baseline, which decomposes to the maximum hierarchy depth using global MGARD kernels, incurs the highest execution time due to repeated full-dataset read-modify-write operations over GPU global memory at each level. Our hybrid hierarchy addresses this by replacing some global levels with In-cache Block levels: under a fixed 3-level decomposition depth case, transitioning from L=1/M=2 to L=2/M=1 consistently reduces execution time across all datasets, achieving a speedup of 1.72$\times$--2.96$\times$ (avg. 2.2$\times$) and 1.82$\times$--3.03$\times$ (avg. 2.3$\times$) over Full Global for decomposition and recomposition respectively, as each substituted global level replaces a full-dataset global memory traversal with on-chip shared memory computation confined within each SM. For applications prioritizing throughput over compression ratio, pure local configurations (L=5, L=3, L=1) remove global levels entirely and further reduce kernel time---reaching a speedup of 2.18$\times$--8.10$\times$ (avg. 4.2$\times$) and 2.78$\times$--6.87$\times$ (avg. 4.2$\times$) over Full Global for decomposition and recomposition respectively.
Hybrid hierarchy with pure local configurations reaches the peak of 8.10$\times$ decomposition and 6.87$\times$ recomposition speedup over Full Global on Hurricane ISABEL, where the dataset's smaller volume makes global memory latency the dominant bottleneck. Across larger datasets such as S3D, speedups remain substantial at 4.05$\times$ and 3.55$\times$, as higher SM concurrency partially amortizes global memory latency even in the baseline. These results demonstrate that the hybrid hierarchy provides a flexible continuum from compression-ratio-preserving configurations to throughput-maximizing ones, all outperforming the full global baseline by exploiting GPU shared memory to eliminate unnecessary global memory traffic.

Figure~\ref{exp_hybrid_decomposition_cr} further examines how the hybrid configuration affects compression ratio relative to full global decomposition. Configurations with a sufficient number of global levels (\emph{e.g.}, $L{=}1, M{=}2$ and $L{=}2, M{=}1$) retain $0.64\times$--$1.78\times$ (avg. 1.7$\times$) of the compression ratio of full global decomposition across NYX, Hurricane, SCALE-LETKF, and Miranda. In contrast, configurations that eliminate global levels entirely (\emph{e.g.}, $L{=}5$, $L{=}3$, $L{=}1$) suffer severe compression ratio degradation across all datasets. S3D is particularly sensitive, dropping to as low as $0.15\times$, owing to its high-frequency content that makes block boundary artifacts more pronounced. These results confirm that retaining at least one global decomposition level is critical for preserving compression quality in the hybrid hierarchy.

\subsection{End-to-End Compression Pipeline}

\begin{table}[t]
\centering
\footnotesize
\setlength{\tabcolsep}{4pt}
\caption{Compression and decompression throughput and compression ratio comparison among MGARD-X (M-X), BlockMGARD (BM), and BlockMGARD-Fused (BMF).}
\label{tab:compression_results}
\resizebox{\columnwidth}{!}{%
\begin{tabular}{@{}ll|rrr|rrr|rr@{}}
\toprule
\multirow{2}{*}{Dataset} & \multirow{2}{*}{Err.} &
\multicolumn{3}{c|}{Comp. (GB/s)} &
\multicolumn{3}{c|}{Decomp. (GB/s)} &
\multicolumn{2}{c@{}}{CR} \\
& & M-X & BM & BMF & M-X & BM & BMF & M-X & BM/BMF \\
\midrule
\multirow{3}{*}{NYX}
& 1e-2 & 33.15 & 55.34 & \textbf{68.36} & 33.65 & 48.79 & \textbf{56.71} & \textbf{29.27} & 26.23 \\
& 1e-4 & 30.24 & 53.54 &  \textbf{65.90} & 19.68 & 48.76 & \textbf{56.41} & 7.52 & \textbf{7.98} \\
& 1e-6 & 27.77 & 50.57 & \textbf{61.15} & 15.96 & 46.52 & \textbf{53.45} & 2.07 & \textbf{2.22} \\
\midrule
\multirow{3}{*}{ISABEL}
& 1e-2 & 10.63 & 30.86 & \textbf{35.71} & 5.32 & 34.51 & \textbf{38.80} & \textbf{19.11} & 18.51 \\
& 1e-4 & 9.01 & 29.16 & \textbf{33.82} & 3.63 & 29.87 & \textbf{33.14} & 4.28 & \textbf{4.37} \\
& 1e-6 & 8.32 & 29.06 & \textbf{34.68} & 3.82 & 30.83 & \textbf{34.15} & 1.13 & \textbf{1.15} \\
\midrule
\multirow{3}{*}{SCALE}
& 1e-2 & 31.62& 51.51&  \textbf{64.49}&   24.34&  45.88&   \textbf{53.34}& \textbf{24.09}& 19.20\\
& 1e-4 & 28.94& 49.75&  \textbf{61.55}&   15.60&  45.03&   \textbf{52.23}&  \textbf{5.59}&  4.69\\
& 1e-6 & 27.63& 47.57&  \textbf{57.88}&   15.04&  45.63&   \textbf{52.86}&  \textbf{1.63}&  1.25 \\
\midrule
\multirow{3}{*}{Miranda}
& 1e-2 & 28.88 & 68.65 & \textbf{82.20} & 19.96 & 68.57 & \textbf{80.42} & \textbf{50.71} & 48.66 \\
& 1e-4 & 23.96 & 65.40 & \textbf{79.95} & 11.61 & 61.92 & \textbf{71.54} & \textbf{20.85} & 20.08 \\
& 1e-6 & 23.15 & 66.52 & \textbf{81.24} & 10.63 & 65.66 & \textbf{76.42} & 3.92 & \textbf{4.69} \\
\midrule
\multirow{3}{*}{S3D}
& 1e-2 & 56.83& 86.28& \textbf{115.71}&   57.80&  81.16&   \textbf{97.82}& \textbf{61.53}& 48.49 \\
& 1e-4 & 54.00& 84.82& \textbf{112.79}&   34.95&  78.34&   \textbf{93.97}& \textbf{23.95}& 17.43 \\
& 1e-6 & 51.36& 83.98& \textbf{110.26}&   29.13&  77.08&   \textbf{92.14}&  \textbf{7.64}&  5.96  \\
\bottomrule
\end{tabular}
}
\end{table}

We compare the compression ratio (CR), compression throughput, and decompression throughput of \ours ($L{=}1, M{=}2$) against MGARD-X across five scientific datasets at error bounds of $1e^{-2}$, $1e^{-4}$, $1e^{-6}$. Results are in Table~\ref{tab:compression_results}, reported for successful compression runs.

\ours consistently outperforms MGARD-X in both compression and decompression throughput, achieving a speedup of 1.52$\times$--3.49$\times$ (avg. 2.2$\times$) for compression and 1.40$\times$--8.23$\times$ (avg. 3.9$\times$) for decompression. In-cache Block decomposition explains the gain: each block touches global memory only twice, avoiding the irregular access patterns that MGARD-X's global decomposition incurs.

However, this speed comes at a cost. Compression ratio drops in several cases, most consistently on SCALE and S3D, since splitting the domain into blocks breaks the spatial locality that global decomposition relies on. This penalty is not uniform, however---on Miranda at $1e^{-6}$, \ours matches or exceeds MGARD-X, suggesting that the effect is dataset-dependent and tolerance-dependent. Users can mitigate the compression ratio loss in practice by increasing $M$ relative to $L$, trading some throughput for better compression ratio.

Kernel fusion keeps the same compression ratio and pushes throughput further---a speedup of 1.16$\times$--1.34$\times$ (avg. 1.2$\times$) for compression and 1.11$\times$--1.21$\times$ (avg. 1.2$\times$) for decompression, on top of \ours. Put together, \ours-Fused reaches a speedup of 2.04$\times$--4.17$\times$ (avg. 2.7$\times$) for compression throughput and 1.69$\times$--9.13$\times$ (avg. 4.6$\times$) for decompression throughput over MGARD-X, with no compression ratio loss over the unfused version. We use this fused configuration as our default for the rest of the experiments.

\subsection{ROI-aware Compression}


\begin{figure*}[t]
\centering
\includegraphics[width=1\linewidth]{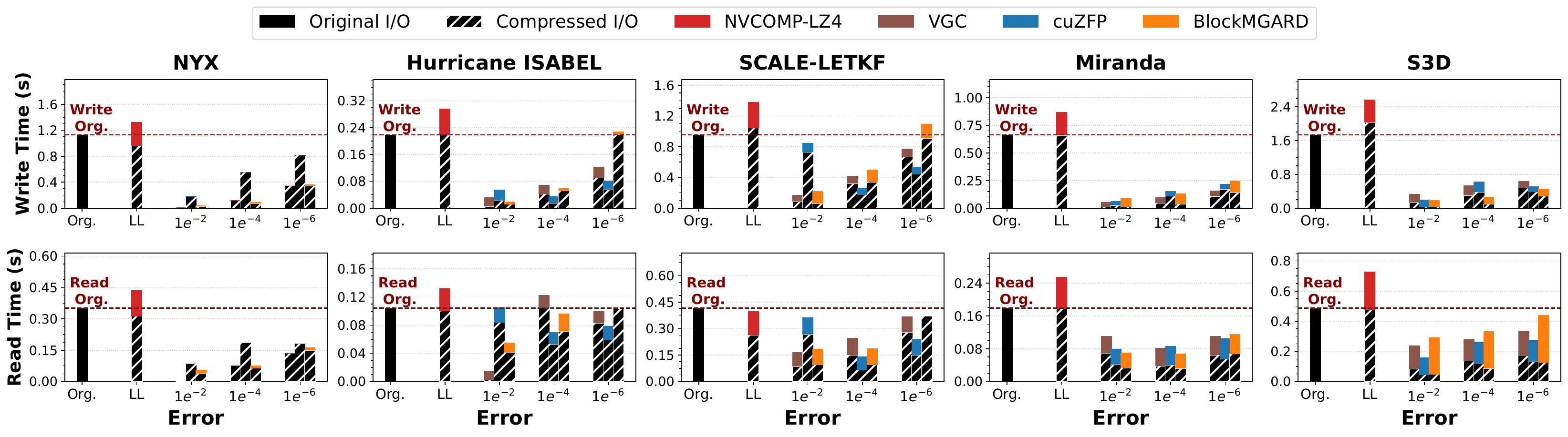}
\caption{Comparing the end-to-end parallel I/O cost of using 4 parallel processes each uses one H100 GPU for (de)compression.}
\label{exp:io}
\end{figure*} 

\paragraph{Higher Compression Ratio at Matched Visual Quality}

We evaluate on the PRES field from SCALE-LETKF, where regions exhibiting large local pressure gradients correspond to active meteorological features such as frontal boundaries and convective systems. We designate a $150 \times 150$ subregion within the $1200 \times 1200$ horizontal slice that encompasses such high-gradient pressure structures as the ROI, assigning a tight relative error tolerance of $1e^{-6}$ to this region while applying a relaxed tolerance of $1e^{-2}$ to the background.

\begin{figure}[htbp]
\centering
\begin{subfigure}[b]{0.24\linewidth}
    \centering
        \includegraphics[width=\linewidth]{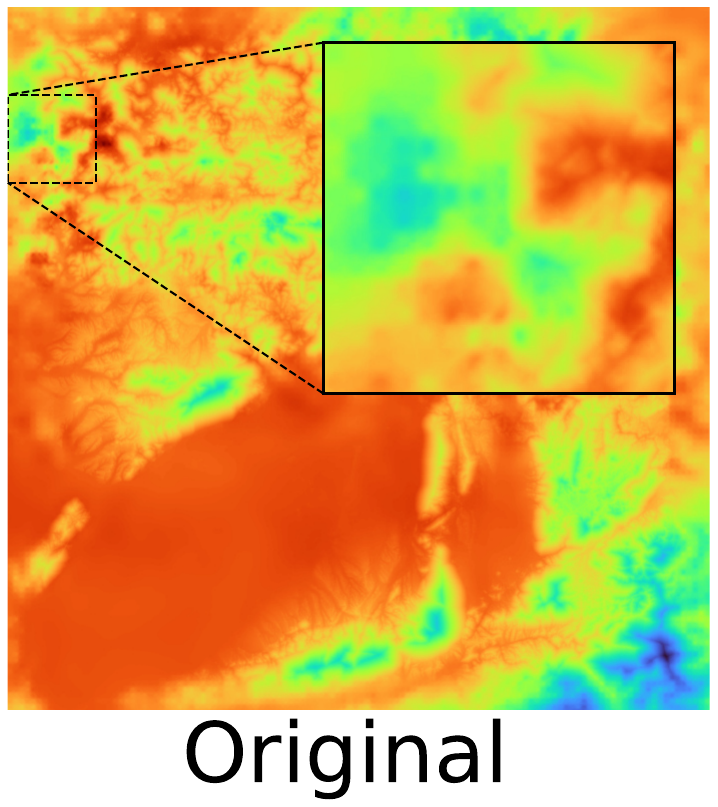}
\end{subfigure}
\hfill
\begin{subfigure}[b]{0.24\linewidth}
    \centering
        \includegraphics[width=\linewidth]{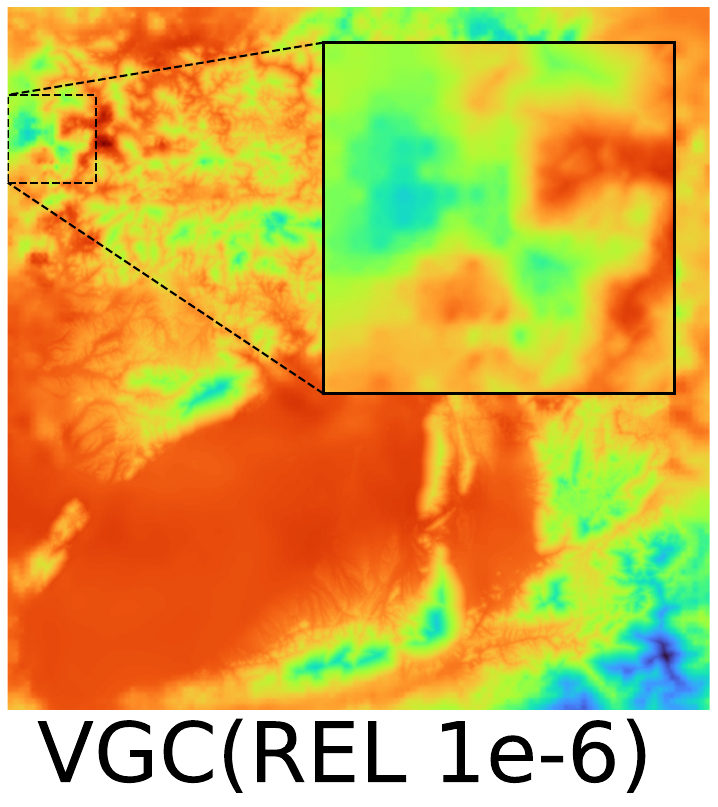}
\end{subfigure}
\hfill
\begin{subfigure}[b]{0.24\linewidth}
    \centering
        \includegraphics[width=\linewidth]{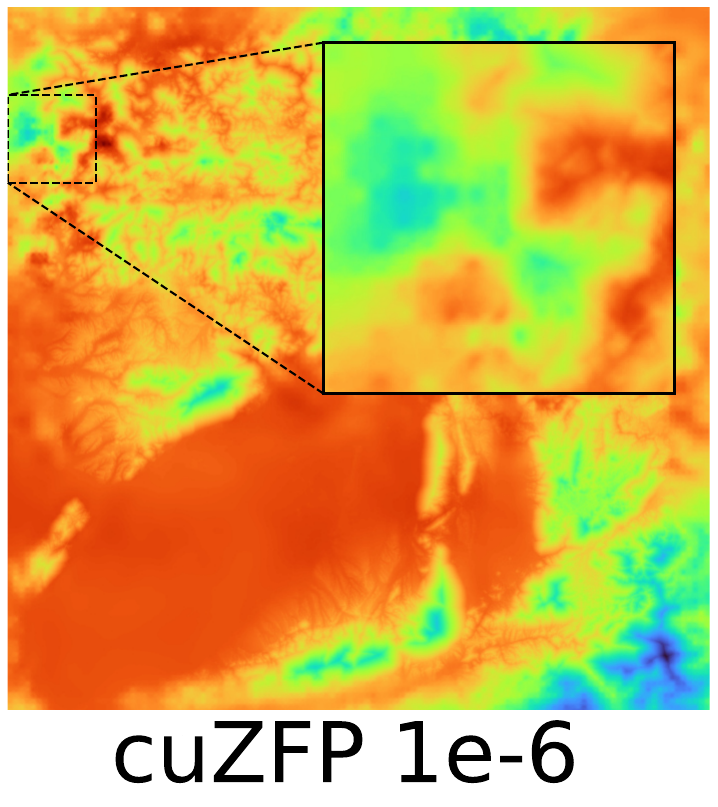}
\end{subfigure}
\hfill
\begin{subfigure}[b]{0.24\linewidth}
    \centering
        \includegraphics[width=\linewidth]{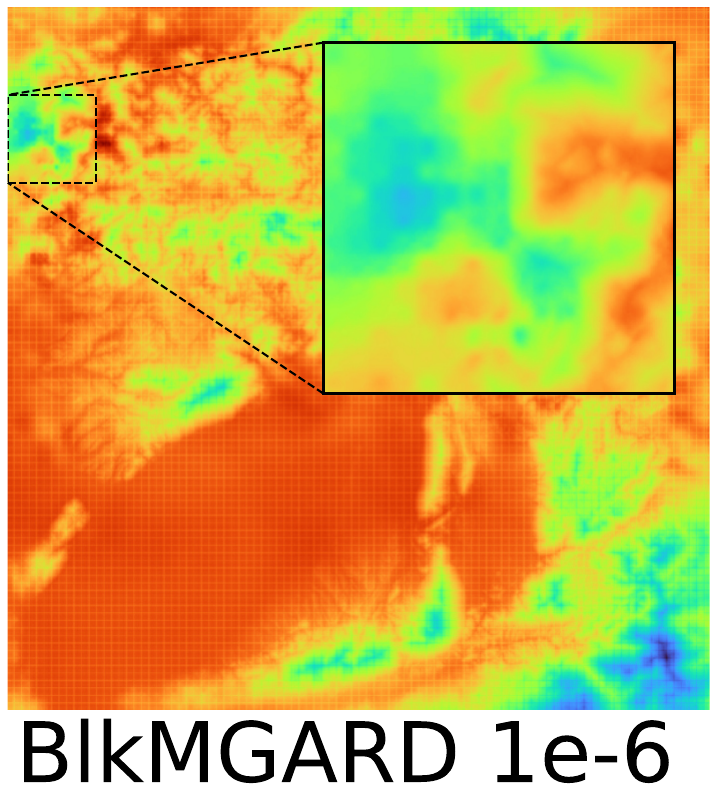}
\end{subfigure}
\hfill
\caption{Visual quality comparison at matched reconstruction fidelity REL $1e^{-6}$. \ours with ROI-aware error control achieves a compression ratio of 5.17$\times$, compared to 3.41$\times$ for cuZFP and 1.67$\times$ for VGC, while preserving spatial features in the ROI (dashed inset). PRES field from SCALE-LETKF, 75th 1200$\times$1200 slice.}
\label{exp_roi_visual_same_quality}
\end{figure}

Figure~\ref{exp_roi_visual_same_quality} presents a visual quality comparison on the 75th slice of the PRES field. VGC and cuZFP apply a uniform tolerance of REL $1e^{-6}$ across the entire domain, achieving compression ratios of 1.67$\times$ and 3.41$\times$ respectively. \ours's ROI-aware compression achieves a compression ratio of 5.17$\times$---a 1.52$\times$ improvement over cuZFP and a 3.10$\times$ improvement over VGC. As highlighted in the zoomed-in insets, the spatial features within the ROI subregion are faithfully reconstructed by BlockMGARD, matching the visual quality of the uniform high-precision baselines within the region of interest. The visible quality degradation in the background regions is an expected and intentional trade-off, reflecting the relaxed tolerance applied outside the ROI. These results demonstrate that \ours's block-granularity error control enables scientists to concentrate reconstruction fidelity in application-critical regions without sacrificing overall compression efficiency.

\begin{figure}[htbp]
\centering
\begin{subfigure}[b]{0.24\linewidth}
    \centering
        \includegraphics[width=\linewidth]{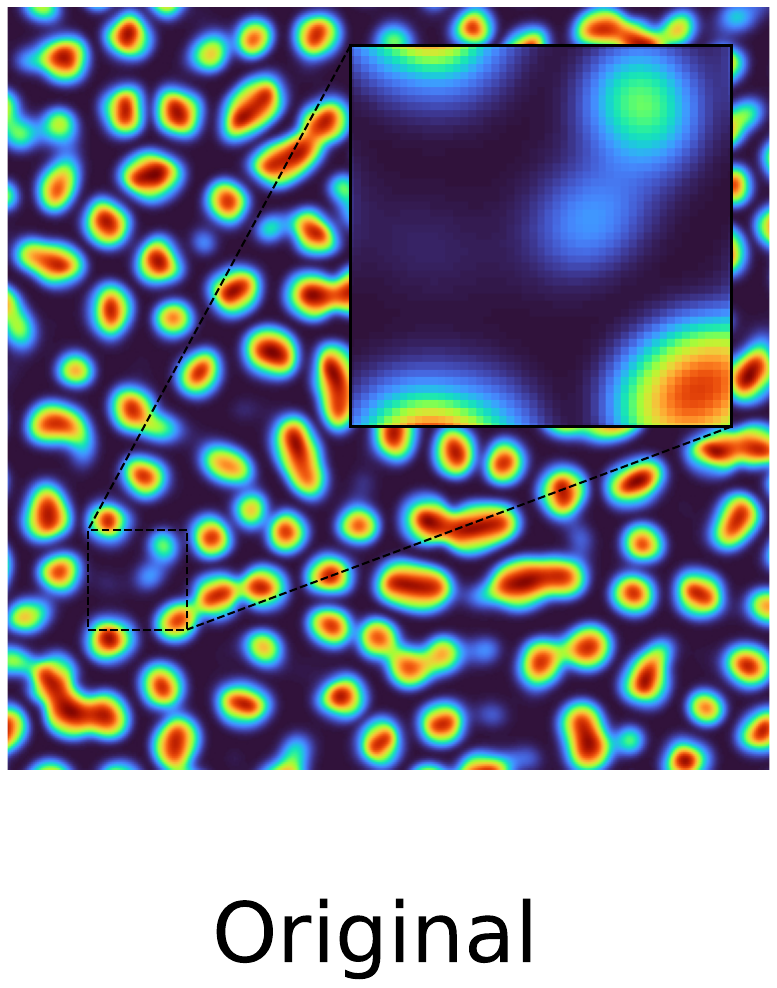}
\end{subfigure}
\hfill
\begin{subfigure}[b]{0.24\linewidth}
    \centering
        \includegraphics[width=\linewidth]{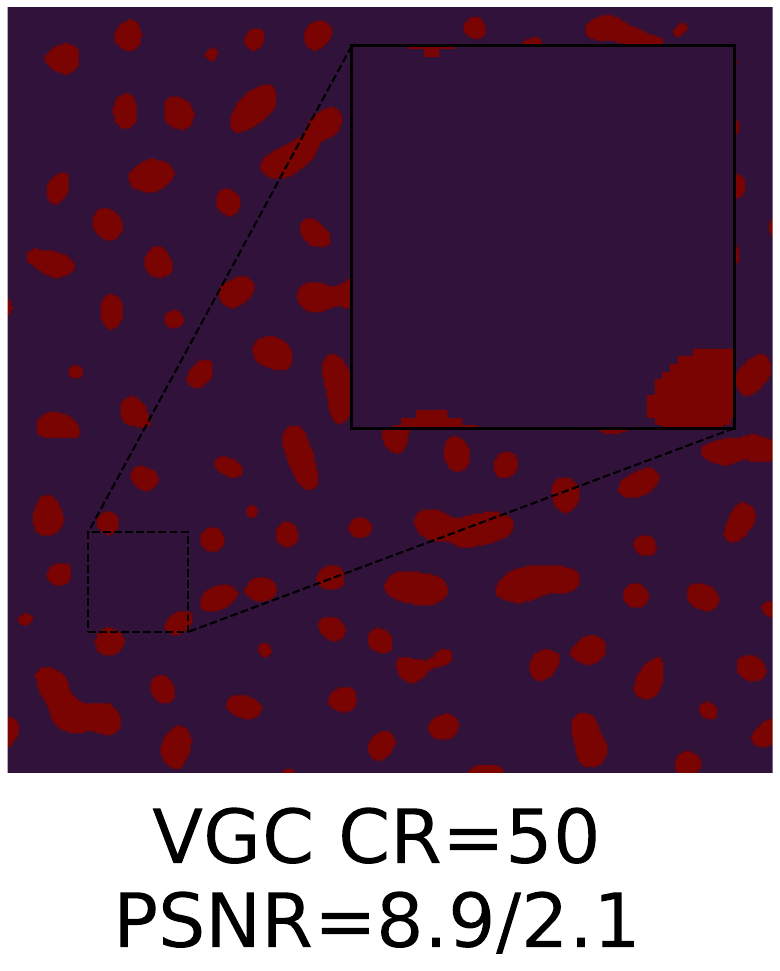}
\end{subfigure}
\hfill
\begin{subfigure}[b]{0.24\linewidth}
    \centering
        \includegraphics[width=\linewidth]{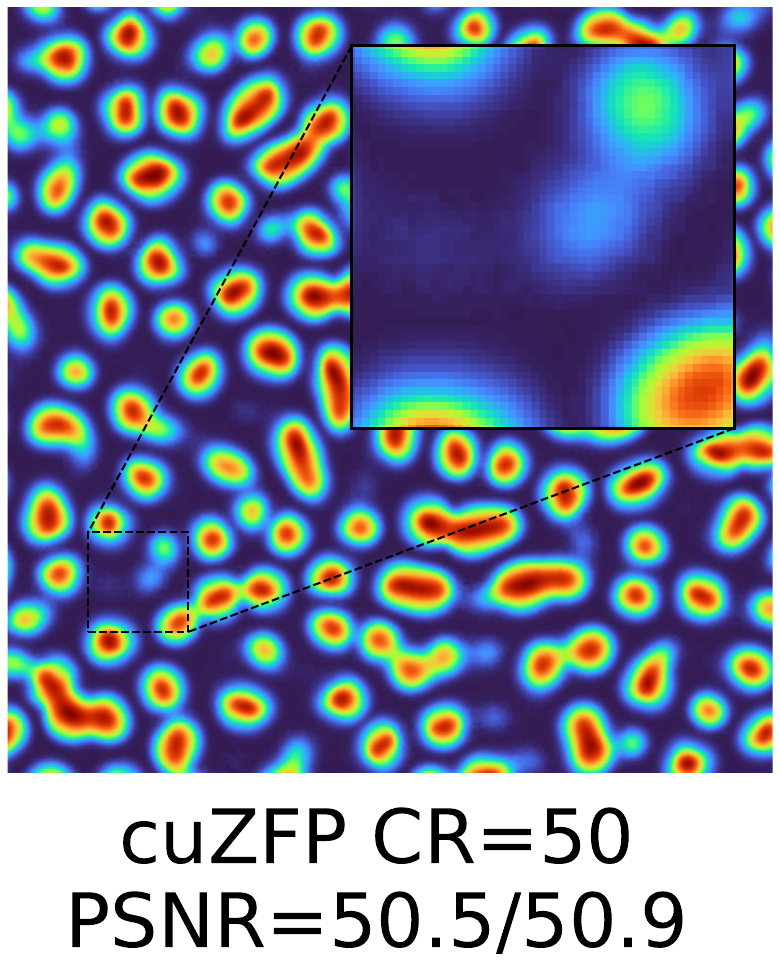}
\end{subfigure}
\hfill
\begin{subfigure}[b]{0.24\linewidth}
    \centering
        \includegraphics[width=\linewidth]{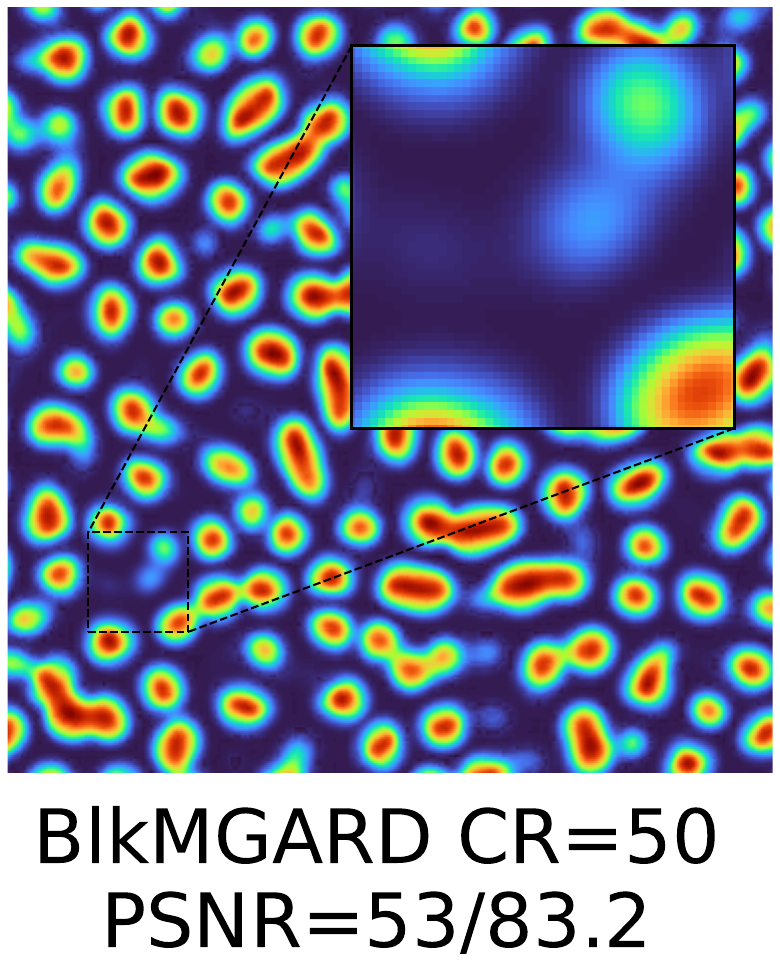}
\end{subfigure}
\caption{Visual fidelity comparison at matched compression ratio ($CR\approx\,50\times$).
PSNR values are reported as global/ROI, where ROI denotes the PSNR within the designated critical region.
Density field from Miranda, 120th $384\times 384$ slice.}
\label{exp_roi_visual_same_cr}
\end{figure}

\paragraph{Superior Visual Fidelity at Matched Compression Ratio}

We further evaluate on the density field from Miranda at a target compression ratio of $50\times$, where all three compressors are configured to produce comparable output sizes. Figure~\ref{exp_roi_visual_same_cr} presents the reconstructed density field at the 120th $384\times384$ slice. At this aggressive compression ratio, the reconstructions reveal stark differences in visual fidelity. VGC suffers severe quality degradation---the zoomed-in inset shows near-complete loss of structural information, with fine-scale density features reduced to a nearly featureless dark region. cuZFP preserves the global structure but introduces visible block-boundary artifacts in the magnified ROI, distorting the smooth gradient transitions characteristic of the original field. In contrast, \ours's ROI-aware compression faithfully preserves both the large-scale structure and fine-grained spatial features within the critical region, with the zoomed inset closely matching the original. These results demonstrate that \ours's block-granularity error control delivers superior reconstruction quality in scientifically sensitive subregions even under aggressive compression, where uniform-tolerance competitors either fail catastrophically or introduce structured artifacts.


\paragraph{Higher Compression Ratio with ROI-Aware Error Control}

Table~\ref{tab:comp-ratio-simple} compares the compression ratios achieved by uniform-tolerance compressors against \ours's ROI-aware mode across five scientific datasets at REL $1e^{-4}$. For uniform compressors, the tolerance is applied globally across the entire domain. For \ours, the same tight tolerance of REL $1e^{-4}$ is enforced within the designated ROI, while a relaxed tolerance is applied to the scientifically less critical background regions---reflecting the realistic scenario where scientists require high fidelity only in application-relevant sub-regions.

\begin{table}[htbp]
    \centering
    \caption{Compression ratio of error-bounded GPU lossy compressors under 1e-4 tolerance. Bold indicates the highest.}
    \label{tab:comp-ratio-simple}
    \begin{tabular}{lcccc}
        \toprule
        \textbf{REL 1E-4} & \textbf{VGC} & \textbf{cuZFP} & \textbf{MGARD-X} & \textbf{BlockMGARD} \\ 
        \midrule
        NYX & 2.29 & 3.42  & 5.13  & \textbf{14.81}  \\
        ISABEL & 2.18 & 3.72  & 3.95  & \textbf{8.99}  \\
        SCALE-LETKF & 6.78 & 3.78  & 3.79  & \textbf{7.89}  \\
        Miranda & 2.88 & 2.92  & 4.03  & \textbf{34.78}  \\
        S3D & 6.30 & 4.93  & 23.28  & \textbf{42.75}  \\
        \bottomrule
    \end{tabular}
\end{table}

Under this setting, \ours achieves substantially higher compression ratios across all datasets. On Miranda and S3D, \ours delivers $34.78\times$ and $42.75\times$ compression ratio respectively---$8.63\times$ and $1.84\times$ higher than the best uniform-tolerance baseline (MGARD-X). On NYX and Hurricane, \ours outperforms the strongest baseline by $2.89\times$ and $2.28\times$ respectively. Even on SCALE-LETKF, where VGC benefits from favorable data characteristics, \ours still achieves a $1.16\times$ improvement. Critically, these gains come without compromising reconstruction fidelity in the ROI: the designated critical subregion is guaranteed the same REL $1e^{-4}$ accuracy as the uniform baselines applied globally, while the overall 
compression efficiency benefits from the relaxed background tolerance. This demonstrates that \ours's block-granularity error control provides a principled mechanism to align compression aggressiveness with scientific data importance.

\begin{figure}[h]
\centering
\includegraphics[width=1\linewidth]{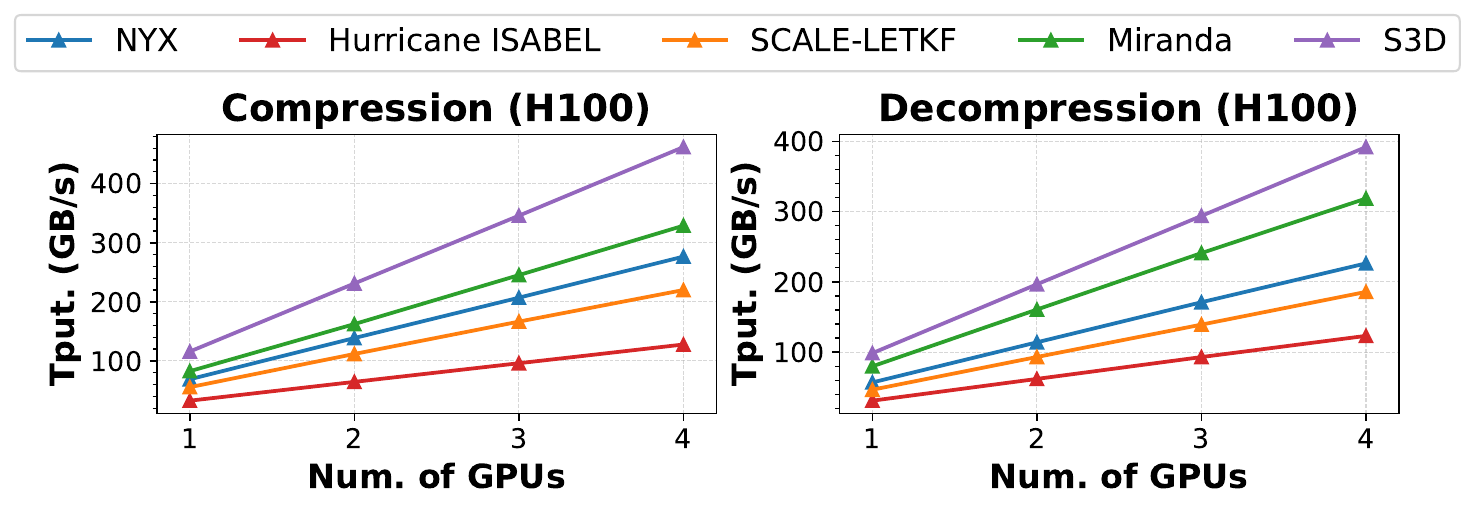}
\caption{Compression and decompression throughput of BlockMGARD scaling up to four H100 GPUs.}
\label{exp:scaling}
\end{figure}

\subsection{Parallel I/O and Scalability}

To evaluate the effectiveness of using data reduction to accelerate parallel I/O at scale~\cite{wang2021unbalanced}, we measure the end-to-end wall time as the sum of compression, write, decompression and read time across all five scientific datasets. Scientific floating-point data exhibits limited compressibility under lossless schemes, resulting in only marginal size reduction. The extra computational overhead consequently increases total I/O time by 1.19--1.48$\times$ over the uncompressed baseline, confirming that lossless compression is ill-suited as an I/O optimization strategy for scientific simulation data.

Among tested GPU lossy compressors, VGC achieves notable acceleration under coarser tolerances, with total I/O speedups of up to 95.4$\times$ on NYX at $1e^{-2}$, benefiting from superior CR of 227. However, its speedup degrades sharply at tighter tolerances (average 3.68$\times$ at $1e^{-4}$, 2.21$\times$ at $1e^{-6}$), as the compression ratio diminishes while the decompression overhead grows. cuZFP delivers more consistent but modest acceleration, ranging from 1.1$\times$--6.1$\times$ at $1e^{-2}$ and 2.0$\times$--3.5$\times$ at $1e^{-4}$, reflecting its fixed-rate compression design.

\ours achieves superior and consistent I/O acceleration by leveraging its high compression ratio across all tolerance levels. At tolerance $1e^{-2}$, \ours reduces total I/O wall time by 3.4$\times$--14.8$\times$ across datasets; at $1e^{-4}$, it still delivers 2.0$\times$--8.6$\times$ speedup. The benefit of write is particularly prominent: the effective write bandwidth improves by up to 25.2$\times$ on NYX, as the GPU compression throughput is sufficient to be fully absorbed within the I/O latency. For Hurricane ISABEL and SCALE-LETKF at tolerance $1e^{-6}$, the total I/O time of \ours approaches the uncompressed baseline ($0.97\times$ and $0.94\times$, respectively). This occurs because these datasets become incompressible at extreme precision requirements, causing \ours to fall back to copying the original data directly and incur compression overhead without any corresponding reduction in data volume.

We additionally evaluate the scalability of \ours by running compression and decompression on up to 4 H100 GPUs within a single node, where each GPU independently processes the same dataset without inter-GPU communication. Figure~\ref{exp:scaling} reports the aggregate throughput across five scientific datasets. \ours achieves near-linear throughput scaling with increasing GPU count, reaching a peak aggregate throughput of 461.9~GB/s for compression and 391.9~GB/s for decompression on S3D dataset at 4 GPUs. The average scaling efficiency across all datasets is 99.3\% for compression and 99.8\% for decompression, demonstrating that \ours introduces negligible multi-GPU overhead.

\section{Conclusion}

In this paper, we present \ours, a GPU-accelerated lossy compressor that redesigns transformation-based data decorrelation to better fit GPU architectures. Specifically, we first introduce In-cache Block decomposition, which confines the entire decomposition working set to shared memory and eliminates runtime index overhead through compile-time lookup tables, achieving up to 5.26$\times$ throughput improvement over global MGARD decomposition. We further propose a hybrid hierarchy that combines local and global decomposition levels, with automatic parameter selection under a user-specified time budget to balance compression speed and ratio. Finally, we develop block-granularity ROI error control with conservative tolerance propagation that guarantees $L_\infty$ error bounds across tolerance boundaries. Experimental evaluation on five real-world scientific datasets demonstrates that \ours consistently outperforms the state-of-the-art MGARD-X lossy compressor in throughput, achieves up to 8.63$\times$ higher compression ratio over uniform-tolerance baselines under ROI-aware mode, and delivers up to 14.8$\times$ I/O cost reduction over uncompressed I/O with near-ideal linear multi-GPU scaling across four H100 GPUs.

\section*{Acknowledgment}
The research is supported in part by the U.S. Department of Energy (DOE) RAPIDS-3 SciDAC and Sirius2 projects under contract number DE-AC05-00OR22725, and the National Science Foundation (NSF) under Grants OAC-2628470, OAC-2628472, OAC-2628473, OAC-2311757, and OAC-2144403. We would like to thank the University of Oregon Research Advanced Computing Services for its support and use of the Talapas clusters.

\bibliographystyle{IEEEtran}
\bibliography{reference}

@techreport{deutsch1996gzip,
  title={GZIP file format specification version 4.3},
  author={Deutsch, Peter},
  year={1996}
}

@article{collet2015zstandard,
  title={Zstandard-real-time data compression algorithm},
  author={Collet, Yann and Kucherawy, EM},
  journal={Zstandard--real-time data compression algorithm},
  year={2015}
}

@ARTICLE{4015488,
  author={Lindstrom, Peter and Isenburg, Martin},
  journal={IEEE Transactions on Visualization and Computer Graphics}, 
  title={Fast and Efficient Compression of Floating-Point Data}, 
  year={2006},
  volume={12},
  number={5},
  pages={1245-1250},
  doi={10.1109/TVCG.2006.143}}

@inproceedings{knorr2021ndzip, author = {Knorr, Fabian and Thoman, Peter and Fahringer, Thomas}, title = {ndzip-gpu: efficient lossless compression of scientific floating-point data on GPUs}, year = {2021}, isbn = {9781450384421}, publisher = {Association for Computing Machinery}, address = {New York, NY, USA}, url = {https://doi.org/10.1145/3458817.3476224}, doi = {10.1145/3458817.3476224}, booktitle = {Proceedings of the International Conference for High Performance Computing, Networking, Storage and Analysis}, articleno = {93}, numpages = {14}, location = {St. Louis, Missouri}, series = {SC '21} }

@INPROCEEDINGS{liang2018error,
  author={Liang, Xin and Di, Sheng and Tao, Dingwen and Li, Sihuan and Li, Shaomeng and Guo, Hanqi and Chen, Zizhong and Cappello, Franck},
  booktitle={2018 IEEE International Conference on Big Data (Big Data)}, 
  title={Error-Controlled Lossy Compression Optimized for High Compression Ratios of Scientific Datasets}, 
  year={2018},
  volume={},
  number={},
  pages={438-447},
  doi={10.1109/BigData.2018.8622520}}

@ARTICLE{liang2022sz3,
  author={Liang, Xin and Zhao, Kai and Di, Sheng and Li, Sihuan and Underwood, Robert and Gok, Ali M. and Tian, Jiannan and Deng, Junjing and Calhoun, Jon C. and Tao, Dingwen and Chen, Zizhong and Cappello, Franck},
  journal={IEEE Transactions on Big Data}, 
  title={SZ3: A Modular Framework for Composing Prediction-Based Error-Bounded Lossy Compressors}, 
  year={2023},
  volume={9},
  number={2},
  pages={485-498},
  doi={10.1109/TBDATA.2022.3201176}}

@INPROCEEDINGS{tao2017significantly,
  author={Tao, Dingwen and Di, Sheng and Chen, Zizhong and Cappello, Franck},
  booktitle={2017 IEEE International Parallel and Distributed Processing Symposium (IPDPS)}, 
  title={Significantly Improving Lossy Compression for Scientific Data Sets Based on Multidimensional Prediction and Error-Controlled Quantization}, 
  year={2017},
  volume={},
  number={},
  pages={1129-1139},
  doi={10.1109/IPDPS.2017.115}}

@INPROCEEDINGS{zhao2021optimizing,
  author={Zhao, Kai and Di, Sheng and Dmitriev, Maxim and Tonellot, Thierry-Laurent D. and Chen, Zizhong and Cappello, Franck},
  booktitle={2021 IEEE 37th International Conference on Data Engineering (ICDE)}, 
  title={Optimizing Error-Bounded Lossy Compression for Scientific Data by Dynamic Spline Interpolation}, 
  year={2021},
  volume={},
  number={},
  pages={1643-1654},
  doi={10.1109/ICDE51399.2021.00145}}

@ARTICLE{lindstrom2014fixed,
  author={Lindstrom, Peter},
  journal={IEEE Transactions on Visualization and Computer Graphics}, 
  title={Fixed-Rate Compressed Floating-Point Arrays}, 
  year={2014},
  volume={20},
  number={12},
  pages={2674-2683},
  doi={10.1109/TVCG.2014.2346458}}

@article{ainsworth2018multilevel,
  title={Multilevel techniques for compression and reduction of scientific data—the univariate case},
  author={Ainsworth, Mark and Tugluk, Ozan and Whitney, Ben and Klasky, Scott},
  journal={Computing and Visualization in Science},
  volume={19},
  number={5},
  pages={65--76},
  year={2018},
  publisher={Springer},
  doi={10.1007/s00791-018-00303-9}
}

@article{ainsworth2019multilevel,
  title={Multilevel techniques for compression and reduction of scientific data---the multivariate case},
  author={Ainsworth, Mark and Tugluk, Ozan and Whitney, Ben and Klasky, Scott},
  journal={SIAM Journal on Scientific Computing},
  volume={41},
  number={2},
  pages={A1278--A1303},
  year={2019},
  publisher={SIAM},
  doi = {10.1137/18M1166651}
}

@article{ainsworth2020multilevel,
author = {Ainsworth, Mark and Tugluk, Ozan and Whitney, Ben and Klasky, Scott},
title = {Multilevel Techniques for Compression and Reduction of Scientific Data---The Unstructured Case},
year = {2020},
issue_date = {2020},
publisher = {Society for Industrial and Applied Mathematics},
address = {USA},
volume = {42},
number = {2},
issn = {1064-8275},
url = {https://doi.org/10.1137/19M1267878},
doi = {10.1137/19M1267878},
journal = {SIAM J. Sci. Comput.},
month = jan,
pages = {A1402–A1427},
numpages = {26}
}

@inproceedings{li2025hp,
author = {Li, Yanliang and Li, Wenbo and Gong, Qian and Liu, Qing and Podhorszki, Norbert and Klasky, Scott and Liang, Xin and Chen, Jieyang},
title = {HP-MDR: High-performance and Portable Data Refactoring and Progressive Retrieval with Advanced GPUs},
year = {2025},
isbn = {9798400714665},
publisher = {Association for Computing Machinery},
address = {New York, NY, USA},
url = {https://doi.org/10.1145/3712285.3759845},
doi = {10.1145/3712285.3759845},
booktitle = {Proceedings of the International Conference for High Performance Computing, Networking, Storage and Analysis},
pages = {2076–2093},
numpages = {18},
location = {
},
series = {SC '25}
}

@inproceedings{tian2020cusz,
author = {Tian, Jiannan and Di, Sheng and Zhao, Kai and Rivera, Cody and Fulp, Megan Hickman and Underwood, Robert and Jin, Sian and Liang, Xin and Calhoun, Jon and Tao, Dingwen and Cappello, Franck},
title = {cuSZ: An Efficient GPU-Based Error-Bounded Lossy Compression Framework for Scientific Data},
year = {2020},
isbn = {9781450380751},
publisher = {Association for Computing Machinery},
address = {New York, NY, USA},
url = {https://doi.org/10.1145/3410463.3414624},
doi = {10.1145/3410463.3414624},
booktitle = {Proceedings of the ACM International Conference on Parallel Architectures and Compilation Techniques},
pages = {3–15},
numpages = {13},
location = {Virtual Event, GA, USA},
series = {PACT '20}
}

@inproceedings{huang2023cuszp,
author = {Huang, Yafan and Di, Sheng and Yu, Xiaodong and Li, Guanpeng and Cappello, Franck},
title = {cuSZp: An Ultra-fast GPU Error-bounded Lossy Compression Framework with Optimized End-to-End Performance},
year = {2023},
isbn = {9798400701092},
publisher = {Association for Computing Machinery},
address = {New York, NY, USA},
url = {https://doi.org/10.1145/3581784.3607048},
doi = {10.1145/3581784.3607048},
booktitle = {Proceedings of the International Conference for High Performance Computing, Networking, Storage and Analysis},
articleno = {43},
numpages = {13},
location = {Denver, CO, USA},
series = {SC '23}
}

@inproceedings{huang2024cuszp2,
author = {Huang, Yafan and Di, Sheng and Li, Guanpeng and Cappello, Franck},
title = {cuSZp2: A GPU Lossy Compressor with Extreme Throughput and Optimized Compression Ratio},
year = {2024},
isbn = {9798350352917},
publisher = {IEEE Press},
url = {https://doi.org/10.1109/SC41406.2024.00021},
doi = {10.1109/SC41406.2024.00021},
booktitle = {Proceedings of the International Conference for High Performance Computing, Networking, Storage, and Analysis},
articleno = {15},
numpages = {18},
location = {Atlanta, GA, USA},
series = {SC '24}
}

@inproceedings{huang2025gpu,
author = {Huang, Yafan and Di, Sheng and Li, Guanpeng and Cappello, Franck},
title = {GPU Lossy Compression for HPC Can Be Versatile and Ultra-Fast},
year = {2025},
isbn = {9798400714665},
publisher = {Association for Computing Machinery},
address = {New York, NY, USA},
url = {https://doi.org/10.1145/3712285.3759817},
doi = {10.1145/3712285.3759817},
booktitle = {Proceedings of the International Conference for High Performance Computing, Networking, Storage and Analysis},
pages = {2056–2075},
numpages = {20},
location = {
},
series = {SC '25}
}

@inproceedings{zhang2025pushing,
author = {Zhang, Boyuan and Huang, Yafan and Di, Sheng and Song, Fengguang and Li, Guanpeng and Cappello, Franck},
title = {Pushing the Limits of GPU Lossy Compression: A Hierarchical Delta Approach},
year = {2025},
isbn = {9798400715372},
publisher = {Association for Computing Machinery},
address = {New York, NY, USA},
url = {https://doi.org/10.1145/3721145.3725743},
doi = {10.1145/3721145.3725743},
booktitle = {Proceedings of the 39th ACM International Conference on Supercomputing},
pages = {654–669},
numpages = {16},
location = {
},
series = {ICS '25}
}

@inproceedings{gong2022region,
author = {Gong, Qian and Whitney, Ben and Zhang, Chengzhu and Liang, Xin and Rangarajan, Anand and Chen, Jieyang and Wan, Lipeng and Ullrich, Paul and Liu, Qing and Jacob, Robert and Ranka, Sanjay and Klasky, Scott},
title = {Region-adaptive, Error-controlled Scientific Data Compression using Multilevel Decomposition},
year = {2022},
isbn = {9781450396677},
publisher = {Association for Computing Machinery},
address = {New York, NY, USA},
url = {https://doi.org/10.1145/3538712.3538717},
doi = {10.1145/3538712.3538717},
booktitle = {Proceedings of the 34th International Conference on Scientific and Statistical Database Management},
articleno = {5},
numpages = {12},
location = {Copenhagen, Denmark},
series = {SSDBM '22}
}

@ARTICLE{cai2019end,
  author={Cai, Chunlei and Chen, Li and Zhang, Xiaoyun and Gao, Zhiyong},
  journal={IEEE Transactions on Image Processing}, 
  title={End-to-End Optimized ROI Image Compression}, 
  year={2020},
  volume={29},
  number={},
  pages={3442-3457},
  doi={10.1109/TIP.2019.2960869}}

@inproceedings{wang2025stz,
author = {Wang, Daoce and Grosset, Pascal and Pulido, Jesus and Tian, Jiannan and Athawale, Tushar and Jia, Jinda and Sun, Baixi and Zhang, Boyuan and Jin, Sian and Zhao, Kai and Ahrens, James and Song, Fengguang},
title = {STZ: A High Quality and High Speed Streaming Lossy Compression Framework for Scientific Data},
year = {2025},
isbn = {9798400714665},
publisher = {Association for Computing Machinery},
address = {New York, NY, USA},
url = {https://doi.org/10.1145/3712285.3759795},
doi = {10.1145/3712285.3759795},
booktitle = {Proceedings of the International Conference for High Performance Computing, Networking, Storage and Analysis},
pages = {2038–2055},
numpages = {18},
location = {
},
series = {SC '25}
}

@INPROCEEDINGS{liang2020toward,
  author={Liang, Xin and Guo, Hanqi and Di, Sheng and Cappello, Franck and Raj, Mukund and Liu, Chunhui and Ono, Kenji and Chen, Zizhong and Peterka, Tom},
  booktitle={2020 IEEE Pacific Visualization Symposium (PacificVis)}, 
  title={Toward Feature-Preserving 2D and 3D Vector Field Compression}, 
  year={2020},
  volume={},
  number={},
  pages={81-90},
  doi={10.1109/PacificVis48177.2020.6431}}

@misc{cuZFP,
  title        = {cuZFP},
  howpublished = {\url{https://github.com/LLNL/zfp/tree/develop/src/cuda_zfp}},
    year={2019}
}

@ARTICLE{jin2025customizable,
  author={Jin, Jian and Xia, Fanxin and Ding, Feng and Zhang, Xinfeng and Liu, Meiqin and Zhao, Yao and Lin, Weisi and Meng, Lili},
  journal={IEEE Transactions on Circuits and Systems for Video Technology}, 
  title={Customizable ROI-Based Deep Image Compression}, 
  year={2026},
  volume={36},
  number={1},
  pages={566-578},
  doi={10.1109/TCSVT.2025.3586792}}

@INPROCEEDINGS {luo2024benchmarking,
author = { Luo, Weile and Fan, Ruibo and Li, Zeyu and Du, Dayou and Wang, Qiang and Chu, Xiaowen },
booktitle = { 2024 IEEE International Parallel and Distributed Processing Symposium (IPDPS) },
title = {{ Benchmarking and Dissecting the Nvidia Hopper GPU Architecture }},
year = {2024},
volume = {},
ISSN = {},
pages = {656-667},
doi = {10.1109/IPDPS57955.2024.00064},
url = {https://doi.ieeecomputersociety.org/10.1109/IPDPS57955.2024.00064},
publisher = {IEEE Computer Society},
address = {Los Alamitos, CA, USA},
month =May}

@INPROCEEDINGS{schroeder2015flying,
  author={Schroeder, William and Maynard, Rob and Geveci, Berk},
  booktitle={2015 IEEE 5th Symposium on Large Data Analysis and Visualization (LDAV)}, 
  title={Flying edges: A high-performance scalable isocontouring algorithm}, 
  year={2015},
  volume={},
  number={},
  pages={33-40},
  doi={10.1109/LDAV.2015.7348069}}

@misc{racs,
  title        = {Research Advanced Computing Services (RACS)},
  howpublished = {\url{racs.uoregon.edu}},
    year={2026}
}

@misc{sdrbench,
  title        = {SDRBench},
  howpublished = {\url{https://sdrbench.github.io}}
}

@INPROCEEDINGS{zhao2020sdrbench,
  author={Zhao, Kai and Di, Sheng and Lian, Xin and Li, Sihuan and Tao, Dingwen and Bessac, Julie and Chen, Zizhong and Cappello, Franck},
  booktitle={2020 IEEE International Conference on Big Data (Big Data)}, 
  title={SDRBench: Scientific Data Reduction Benchmark for Lossy Compressors}, 
  year={2020},
  volume={},
  number={},
  pages={2716-2724},
  doi={10.1109/BigData50022.2020.9378449}}

@article{almgren2013nyx,
doi = {10.1088/0004-637X/765/1/39},
url = {https://doi.org/10.1088/0004-637X/765/1/39},
year = {2013},
month = {feb},
publisher = {The American Astronomical Society},
volume = {765},
number = {1},
pages = {39},
author = {Almgren, Ann S. and Bell, John B. and Lijewski, Mike J. and Lukić, Zarija and Van Andel, Ethan},
title = {Nyx: A MASSIVELY PARALLEL AMR CODE FOR COMPUTATIONAL COSMOLOGY},
journal = {The Astrophysical Journal}
}

@article{cook2004mixing, title={The mixing transition in Rayleigh–Taylor instability}, volume={511}, DOI={10.1017/S0022112004009681}, journal={Journal of Fluid Mechanics}, author={COOK, ANDREW W. and CABOT, WILLIAM and MILLER, PAUL L.}, year={2004}, pages={333–362}}

@misc{hurricane,
  title        = {Hurricane ISABEL Simulation Data},
  howpublished = {\url{http://vis.computer.org/vis2004contest/data.html}},
  year={2019}
}

@article{hunt2007efficient,
title = {Efficient data assimilation for spatiotemporal chaos: A local ensemble transform Kalman filter},
journal = {Physica D: Nonlinear Phenomena},
volume = {230},
number = {1},
pages = {112-126},
year = {2007},
note = {Data Assimilation},
issn = {0167-2789},
doi = {https://doi.org/10.1016/j.physd.2006.11.008},
url = {https://www.sciencedirect.com/science/article/pii/S0167278906004647},
author = {Brian R. Hunt and Eric J. Kostelich and Istvan Szunyogh}
}

@article{chen2009terascale,
doi = {10.1088/1749-4699/2/1/015001},
url = {https://doi.org/10.1088/1749-4699/2/1/015001},
year = {2009},
month = {jan},
publisher = {},
volume = {2},
number = {1},
pages = {015001},
author = {Chen, J H and Choudhary, A and de Supinski, B and DeVries, M and Hawkes, E R and Klasky, S and Liao, W K and Ma, K L and Mellor-Crummey, J and Podhorszki, N and Sankaran, R and Shende, S and Yoo, C S},
title = {Terascale direct numerical simulations of turbulent combustion using S3D},
journal = {Computational Science \& Discovery}
}

@article{dongarra2011international,
author = {Dongarra, Jack and Beckman, Pete and Moore, Terry and Aerts, Patrick and Aloisio, Giovanni and Andre, Jean-Claude and Barkai, David and Berthou, Jean-Yves and Boku, Taisuke and Braunschweig, Bertrand and Cappello, Franck and Chapman, Barbara and Xuebin Chi and Choudhary, Alok and Dosanjh, Sudip and Dunning, Thom and Fiore, Sandro and Geist, Al and Gropp, Bill and Harrison, Robert and Hereld, Mark and Heroux, Michael and Hoisie, Adolfy and Hotta, Koh and Zhong Jin and Ishikawa, Yutaka and Johnson, Fred and Kale, Sanjay and Kenway, Richard and Keyes, David and Kramer, Bill and Labarta, Jesus and Lichnewsky, Alain and Lippert, Thomas and Lucas, Bob and Maccabe, Barney and Matsuoka, Satoshi and Messina, Paul and Michielse, Peter and Mohr, Bernd and Mueller, Matthias S. and Nagel, Wolfgang E. and Nakashima, Hiroshi and Papka, Michael E and Reed, Dan and Sato, Mitsuhisa and Seidel, Ed and Shalf, John and Skinner, David and Snir, Marc and Sterling, Thomas and Stevens, Rick and Streitz, Fred and Sugar, Bob and Sumimoto, Shinji and Tang, William and Taylor, John and Thakur, Rajeev and Trefethen, Anne and Valero, Mateo and Van Der Steen, Aad and Vetter, Jeffrey and Williams, Peg and Wisniewski, Robert and Yelick, Kathy},
title = {The International Exascale Software Project roadmap},
year = {2011},
issue_date = {February  2011},
publisher = {Sage Publications, Inc.},
address = {USA},
volume = {25},
number = {1},
issn = {1094-3420},
url = {https://doi.org/10.1177/1094342010391989},
doi = {10.1177/1094342010391989},
journal = {Int. J. High Perform. Comput. Appl.},
month = feb,
pages = {3–60},
numpages = {58}
}

@article{io_bottleneck_2014,
author = {Kogge, Peter and Borkar, S. and Campbell, Dan and Carlson, William and Dally, William and Denneau, Monty and Franzon, Paul and Harrod, William and Hiller, Jon and Keckler, Stephen and Klein, Dean and Lucas, Robert},
year = {2008},
month = {01},
pages = {},
title = {ExaScale Computing Study: Technology Challenges in Achieving Exascale Systems},
volume = {15},
journal = {Defense Advanced Research Projects Agency Information Processing Techniques Office (DARPA IPTO), Techinal Representative},
url ={https://ftp.eecs.berkeley.edu/~yelick/papers/Exascale_final_report.pdf}
}

@article{chang2009compressed,
  title={Compressed ion temperature gradient turbulence in diverted tokamak plasmas},
  author={Chang, C. S. and Ku, S. and Weitzner, H.},
  journal={Physics of Plasmas},
  volume={16},
  number={5},
  pages={056108},
  year={2009},
  doi = {10.1063/1.3099329}
}

@article{ku2018full,
  title={Full-f gyrokinetic particle simulation of tokamak plasma turbulence in realistic geometry},
  author={Ku, S. and Chang, C. S. and Hager, R. and others},
  journal={Physics of Plasmas},
  volume={25},
  number={5},
  pages={056101},
  year={2018},
  doi = {10.1088/0029-5515/49/11/115021}
}

@article{vay2018warpx,
  title={WarpX: A New Exascale Computing Platform for Beam--Plasma Simulations},
  author={Vay, Jean-Luc and Almgren, Ann and Bell, John and others},
  journal={Plasma Physics and Controlled Fusion},
  volume={60},
  number={1},
  pages={014001},
  year={2018},
  doi = {10.1016/j.nima.2018.01.035}
}

@ARTICLE{burau2010picongpu,
  author={Burau, Heiko and Widera, Renée and Hönig, Wolfgang and Juckeland, Guido and Debus, Alexander and Kluge, Thomas and Schramm, Ulrich and Cowan, Tomas E. and Sauerbrey, Roland and Bussmann, Michael},
  journal={IEEE Transactions on Plasma Science}, 
  title={PIConGPU: A Fully Relativistic Particle-in-Cell Code for a GPU Cluster}, 
  year={2010},
  volume={38},
  number={10},
  pages={2831-2839},
  doi={10.1109/TPS.2010.2064310}}

@INPROCEEDINGS{8888528,
  author={Labate, Maria Grazia and Waterson, Mark and Swart, Gerhard and Bowen, Mark and Dewdney, Peter},
  booktitle={2019 IEEE International Symposium on Antennas and Propagation and USNC-URSI Radio Science Meeting}, 
  title={The Square Kilometre Array Observatory}, 
  year={2019},
  volume={},
  number={},
  pages={391-392},
  doi={10.1109/APUSNCURSINRSM.2019.8888528}}

@article{dodson2024optimising, title={Optimising the processing and storage of visibilities using lossy compression}, volume={42}, DOI={10.1017/pasa.2025.29}, journal={Publications of the Astronomical Society of Australia}, author={Dodson, Richard and Williamson, Alexander and Gong, Qian and Elahi, Pascal and Wicenec, Andreas and Rioja, María J. and Chen, Jieyang and Podhorszki, Norbert and Klasky, Scott and Meyer, Martin}, year={2025}, pages={e093}}

@article{gong2023mgard,
  title={MGARD: A multigrid framework for high-performance, error-controlled data compression and refactoring},
  author={Gong, Qian and Chen, Jieyang and Whitney, Ben and Liang, Xin and Reshniak, Viktor and Banerjee, Tania and Lee, Jaemoon and Rangarajan, Anand and Wan, Lipeng and Vidal, Nicolas and others},
  journal={SoftwareX},
  volume={24},
  pages={101590},
  year={2023},
  publisher={Elsevier},
  doi = {10.1016/j.softx.2023.101590}
}

@article{liang2021mgard+,
  title={Mgard+: Optimizing multilevel methods for error-bounded scientific data reduction},
  author={Liang, Xin and Whitney, Ben and Chen, Jieyang and Wan, Lipeng and Liu, Qing and Tao, Dingwen and Kress, James and Pugmire, David and Wolf, Matthew and Podhorszki, Norbert and others},
  journal={IEEE Transactions on Computers},
  volume={71},
  number={7},
  pages={1522--1536},
  year={2021},
  publisher={IEEE},
  doi = {10.1109/TC.2021.3092201}
}

@article{ainsworth2019qoi,
  title={Multilevel techniques for compression and reduction of scientific data-quantitative control of accuracy in derived quantities},
  author={Ainsworth, Mark and Tugluk, Ozan and Whitney, Ben and Klasky, Scott},
  journal={SIAM Journal on Scientific Computing},
  volume={41},
  number={4},
  pages={A2146--A2171},
  year={2019},
  publisher={SIAM},
  doi = {/10.1137/18M1208885}
}

@article{lindstrom2025zfp,
  title={ZFP: A compressed array representation for numerical computations},
  author={Lindstrom, Peter and Hittinger, Jeffrey and Diffenderfer, James and Fox, Alyson and Osei-Kuffuor, Daniel and Banks, Jeffrey},
  journal={The International Journal of High Performance Computing Applications},
  volume={39},
  number={1},
  pages={104--122},
  year={2025},
  publisher={SAGE Publications Sage UK: London, England},
  doi = {10.1177/10943420241284023}
}

@ARTICLE{8031063,
  author={Di, Sheng and Cappello, Franck},
  journal={IEEE Transactions on Parallel and Distributed Systems}, 
  title={Optimization of Error-Bounded Lossy Compression for Hard-to-Compress HPC Data}, 
  year={2018},
  volume={29},
  number={1},
  pages={129-143},
  doi={10.1109/TPDS.2017.2749300}}

@INPROCEEDINGS{7967203,
  author={Tao, Dingwen and Di, Sheng and Chen, Zizhong and Cappello, Franck},
  booktitle={2017 IEEE International Parallel and Distributed Processing Symposium (IPDPS)}, 
  title={Significantly Improving Lossy Compression for Scientific Data Sets Based on Multidimensional Prediction and Error-Controlled Quantization}, 
  year={2017},
  volume={},
  number={},
  pages={1129-1139},
  doi={10.1109/IPDPS.2017.115}}

@INPROCEEDINGS{chen2021accelerating,
  author={Chen, Jieyang and Wan, Lipeng and Liang, Xin and Whitney, Ben and Liu, Qing and Pugmire, David and Thompson, Nicholas and Choi, Jong Youl and Wolf, Matthew and Munson, Todd and Foster, Ian and Klasky, Scott},
  booktitle={2021 IEEE International Parallel and Distributed Processing Symposium (IPDPS)}, 
  title={Accelerating Multigrid-based Hierarchical Scientific Data Refactoring on GPUs}, 
  year={2021},
  volume={},
  number={},
  pages={859-868},
  doi={10.1109/IPDPS49936.2021.00095}}

@INPROCEEDINGS{8048933,
  author={Li, Shaomeng and Sane, Sudhanshu and Orf, Leigh and Mininni, Pablo and Clyne, John and Childs, Hank},
  booktitle={2017 IEEE International Conference on Cluster Computing (CLUSTER)}, 
  title={Spatiotemporal Wavelet Compression for Visualization of Scientific Simulation Data}, 
  year={2017},
  volume={},
  number={},
  pages={216-227},
  doi={10.1109/CLUSTER.2017.15}}

@INPROCEEDINGS{jin2022accelerating,
  author={Jin, Sian and Tao, Dingwen and Tang, Houjun and Di, Sheng and Byna, Suren and Lukic, Zarija and Cappello, Franck},
  booktitle={SC22: International Conference for High Performance Computing, Networking, Storage and Analysis}, 
  title={Accelerating Parallel Write via Deeply Integrating Predictive Lossy Compression with HDF5}, 
  year={2022},
  volume={},
  number={},
  pages={1-15},
  doi={10.1109/SC41404.2022.00066}}

@INPROCEEDINGS{chen2025hpdr,
  author={Chen, Jieyang and Gong, Qian and Li, Yanliang and Liang, Xin and Wan, Lipeng and Liu, Qing and Podhorszki, Norbert and Klasky, Scott},
  booktitle={2025 IEEE International Parallel and Distributed Processing Symposium (IPDPS)}, 
  title={HPDR: High-Performance Portable Scientific Data Reduction Framework}, 
  year={2025},
  volume={},
  number={},
  pages={1104-1116},
  doi={10.1109/IPDPS64566.2025.00101}}

@INPROCEEDINGS{lu2023zfp,
  author={Lu, Bing and Li, Yida and Wang, Junqi and Luo, Huizhang and Li, Kenli},
  booktitle={2023 IEEE International Parallel and Distributed Processing Symposium (IPDPS)}, 
  title={ZFP-X: Efficient Embedded Coding for Accelerating Lossy Floating Point Compression}, 
  year={2023},
  volume={},
  number={},
  pages={1041-1050},
  doi={10.1109/IPDPS54959.2023.00107}}

@phdthesis{Volkov:EECS-2016-143,
    Author= {Volkov, Vasily},
    Title= {Understanding Latency Hiding on GPUs},
    School= {EECS Department, University of California, Berkeley},
    Year= {2016},
    Month= {Aug},
    Url= {http://www2.eecs.berkeley.edu/Pubs/TechRpts/2016/EECS-2016-143.html},
    Number= {UCB/EECS-2016-143},
}

@article{little1961proof,
  title={A proof for the queuing formula: L= $\lambda$ W},
  author={Little, John DC},
  journal={Operations research},
  volume={9},
  number={3},
  pages={383--387},
  year={1961},
  publisher={INFORMS},
  doi = {doi.org/10.1287/opre.9.3.383}
}

@INPROCEEDINGS{li2024accelerating,
  author={Li, Yanliang and Chen, Jieyang},
  booktitle={2024 IEEE 20th International Conference on e-Science (e-Science)}, 
  title={Accelerating In-transit Isosurface Generation With Topology Preserving Compression}, 
  year={2024},
  volume={},
  number={},
  pages={1-3},
  doi={10.1109/e-Science62913.2024.10678711}}

@article{Bilderback_2005,
doi = {10.1088/0953-4075/38/9/022},
url = {https://doi.org/10.1088/0953-4075/38/9/022},
year = {2005},
month = {apr},
publisher = {},
volume = {38},
number = {9},
pages = {S773},
author = {Bilderback, Donald H and Elleaume, Pascal and Weckert, Edgar},
title = {Review of third and next generation synchrotron light sources},
journal = {Journal of Physics B: Atomic, Molecular and Optical Physics}
}

@INPROCEEDINGS{tian2021huffman,
  author={Tian, Jiannan and Rivera, Cody and Di, Sheng and Chen, Jieyang and Liang, Xin and Tao, Dingwen and Cappello, Franck},
  booktitle={2021 IEEE International Parallel and Distributed Processing Symposium (IPDPS)},
  title={Revisiting Huffman Coding: Toward Extreme Performance on Modern GPU Architectures},
  year={2021},
  pages={881-891},
  doi={10.1109/IPDPS49936.2021.00097}
}

@INPROCEEDINGS{wang2021unbalanced,
  author={Wang, Xinying and Wan, Lipeng and Chen, Jieyang and Gong, Qian and Whitney, Ben and Wang, Jinzhen and Gainaru, Ana and Liu, Qing and Podhorszki, Norbert and Zhao, Dongfang and Yan, Feng and Klasky, Scott},
  booktitle={2021 7th International Workshop on Data Analysis and Reduction for Big Scientific Data (DRBSD)},
  title={Unbalanced Parallel {I/O}: An Often-Neglected Side Effect of Lossy Scientific Data Compression},
  year={2021},
  doi={10.1109/DRBSD754563.2021.00008}
}

@INPROCEEDINGS{yakushin2020feature,
  author={Yakushin, Igor and Mehta, Kshitij and Chen, Jieyang and Wolf, Matthew and Foster, Ian and Klasky, Scott and Munson, Todd},
  booktitle={Workshop Proceedings of the 49th International Conference on Parallel Processing (ICPP Workshops)},
  title={Feature-preserving Lossy Compression for In Situ Data Analysis},
  year={2020},
  doi={10.1145/3409390.3409400}
}

@INPROCEEDINGS{gong2023spatiotemporal,
  author={Gong, Qian and Zhang, Chengzhu and Liang, Xin and Reshniak, Viktor and Chen, Jieyang and Rangarajan, Anand and Ranka, Sanjay and Vidal, Nicolas and Wan, Lipeng and Ullrich, Paul and Podhorszki, Norbert and Jacob, Robert and Klasky, Scott},
  booktitle={2023 IEEE 19th International Conference on e-Science (e-Science)},
  title={Spatiotemporally Adaptive Compression for Scientific Dataset with Feature Preservation---A Case Study on Simulation Data with Extreme Climate Events Analysis},
  year={2023}
}

@INPROCEEDINGS{gong2022trust,
  author={Gong, Qian and Liang, Xin and Whitney, Ben and Choi, Jong Youl and Chen, Jieyang and Wan, Lipeng and Ethier, Stephane and Ku, Seung-Hoe and Churchill, Michael and Chang, Choong-Seock and Ainsworth, Mark and Tugluk, Ozan and Munson, Todd and Pugmire, Dave and Archibald, Rick and Klasky, Scott},
  booktitle={Smoky Mountains Computational Sciences and Engineering Conference (SMC)},
  title={Maintaining Trust in Reduction: Preserving the Accuracy of Quantities of Interest for Lossy Compression},
  year={2022},
  doi={10.1007/978-3-030-96498-6_2}
}

@INPROCEEDINGS{banerjee2023online,
  author={Banerjee, Tania and Lee, Jaemoon and Choi, Jong and Gong, Qian and Chen, Jieyang and Chang, Choong-Seock and Klasky, Scott and Rangarajan, Anand and Ranka, Sanjay},
  booktitle={2023 IEEE 19th International Conference on e-Science (e-Science)},
  title={Online and Scalable Data Compression Pipeline with Guarantees on Quantities of Interest},
  year={2023}
}

@INPROCEEDINGS{chen2019understanding,
  author={Chen, Jieyang and Pugmire, Dave and Wolf, Matthew and Thompson, Nick and Logan, Jeremy and Mehta, Kshitij and Wan, Lipeng and Choi, Jong Youl and Whitney, Ben and Klasky, Scott},
  booktitle={2019 IEEE/ACM 5th International Workshop on Data Analysis and Reduction for Big Scientific Data (DRBSD)},
  title={Understanding Performance-Quality Trade-offs in Scientific Visualization Workflows with Lossy Compression},
  year={2019}
}

@ARTICLE{wan2022improving,
  author={Wan, Lipeng and Huebl, Axel and Gu, Junmin and Poeschel, Franz and Gainaru, Ana and Wang, Ruonan and Chen, Jieyang and Liang, Xin and Ganyushin, Dmitry and Munson, Todd and Foster, Ian and Vay, Jean-Luc and Podhorszki, Norbert and Wu, Kesheng and Klasky, Scott},
  journal={IEEE Transactions on Parallel and Distributed Systems},
  title={Improving {I/O} Performance for Exascale Applications Through Online Data Layout Reorganization},
  year={2022},
  volume={33},
  number={4},
  pages={878-890},
  doi={10.1109/TPDS.2021.3100784}
}

@INPROCEEDINGS{gong2024general,
  author={Gong, Qian and Wang, Zhe and Reshniak, Viktor and Liang, Xin and Chen, Jieyang and Liu, Qing and Athawale, Tushar M. and Ju, Yi and Rangarajan, Anand and Ranka, Sanjay and Podhorszki, Norbert and Archibald, Rick and Klasky, Scott},
  booktitle={2024 IEEE 20th International Conference on e-Science (e-Science)},
  title={A General Framework for Error-controlled Unstructured Scientific Data Compression},
  year={2024}
}

\end{document}